\documentclass[prd,twocolumn,showpacs,amsmath,amssymb]{revtex4-2}

\usepackage{subfigure}
\usepackage{enumerate}
\usepackage{graphicx}
\usepackage{epstopdf}
\usepackage{soul}
\usepackage{dcolumn}
\usepackage{color}
\usepackage{bm}
\usepackage{overpic}
\usepackage{rotating}
\usepackage{times}
\usepackage{multirow}
\usepackage{float}
\usepackage{mathrsfs}
\usepackage{booktabs}
\usepackage{float}
\usepackage[T1]{fontenc}
\usepackage{xcolor}
\usepackage[colorlinks,linkcolor=blue,anchorcolor=blue,citecolor=blue]{hyperref}

\newcommand{\BR}{{\cal B}}

\newcommand{\pip}{\pi^+}
\newcommand{\pim}{\pi^-}

\newcommand{\etap}{\eta^\prime}

\newcommand{\jpsi}{J/\psi}

\newcommand{\etamp}{\eta_{1}(1855)}

\newcommand{\pimp}{\pi^+\pi^-}
\newcommand{\pio}{\pi^0}

\begin{document}

%\linenumbers
%\DeclareGraphicsExtensions{.eps,.png,.ps}
% The following two lines are used to deal with the hyphenation problem.
%\hyphenpenalty=10000
\lefthyphenmin=2
\righthyphenmin=2
\tolerance=1000
\uchyph=0

\normalsize
\parskip=5pt plus 1pt minus 1pt
%\pagewiselinenumbers
%\title{Helicity Amplitude Analysis of $\psip\to \OOb$}
%\title{\boldmath Observation of isoscalar an $1^{-+}$ resonance in $\jpsi\rightarrow\gamma\eta\eta'$}
\title{\boldmath Partial wave analysis of $\jpsi\rightarrow\gamma\eta\eta'$}
\author{
M.~Ablikim$^{1}$, M.~N.~Achasov$^{10,b}$, P.~Adlarson$^{68}$, S. ~Ahmed$^{14}$, M.~Albrecht$^{4}$, R.~Aliberti$^{28}$, A.~Amoroso$^{67A,67C}$, M.~R.~An$^{32}$, Q.~An$^{64,50}$, X.~H.~Bai$^{58}$
% ...
    \\
        \vspace{0.2cm}
        (BESIII Collaboration)\\
        \vspace{0.2cm} {\it
$^{1}$ Institute of High Energy Physics, Beijing 100049, People's Republic of China\\
% ...
$^{4}$ Bochum Ruhr-University, D-44780 Bochum, Germany\\
% ...
$^{10}$ G.I. Budker Institute of Nuclear Physics SB RAS (BINP), Novosibirsk 630090, Russia\\
% ...
$^{14}$ Helmholtz Institute Mainz, Staudinger Weg 18, D-55099 Mainz, Germany\\
% ...
$^{28}$ Johannes Gutenberg University of Mainz, Johann-Joachim-Becher-Weg 45, D-55099 Mainz, Germany\\
% ...
$^{32}$ Liaoning Normal University, Dalian 116029, People's Republic of China\\
% ...
$^{50}$ State Key Laboratory of Particle Detection and Electronics, Beijing 100049, Hefei 230026, People's Republic of China\\
% ...
$^{58}$ University of Jinan, Jinan 250022, People's Republic of China\\
% ...
$^{64}$ University of Science and Technology of China, Hefei 230026, People's Republic of China\\
% ...
$^{67}$ University of Turin and INFN, (A)University of Turin, I-10125, Turin, Italy; (B)University of Eastern Piedmont, I-15121, Alessandria, Italy; (C)INFN, I-10125, Turin, Italy\\
$^{68}$ Uppsala University, Box 516, SE-75120 Uppsala, Sweden\\
% ...
$^{b}$ Also at the Novosibirsk State University, Novosibirsk, 630090, Russia\\
% ...
        }
}

\date{\today}

\begin{abstract}
Based on a sample of (10.09$\pm$0.04)$\times$10$^{9}$ $\jpsi$ events collected with the BESIII detector operating at the BEPCII storage ring, a partial wave analysis of the decay $\jpsi \rightarrow \gamma\eta\etap$ is performed. An isoscalar state with exotic quantum numbers $J^{PC}=1^{-+}$, denoted as $\etamp$, has been observed for the first time with statistical significance larger than 19$\sigma$. Its mass and width are measured to be (1855$\pm$9$_{-1}^{+6}$)~MeV/$c^{2}$ and (188$\pm$18$_{-8}^{+3}$)~MeV, respectively. The first uncertainties are statistical and the second are systematic. The product branching fraction  $\BR(J/\psi$$\rightarrow$$ \gamma\eta_1(1855))$$\BR(\eta_1(1855)\rightarrow$$\eta\eta')$ is measured to be (2.70$\pm  0.41 _{-0.35}^{+0.16}) \times$10$^{-6}$.
 In addition, an upper limit on the ratio of branching fractions $\BR(f_0(1710)$$\rightarrow$$\eta\etap)$/$\BR(f_0(1710)$$\rightarrow$$\pi\pi)$ is determined to be $2.87 \times 10^{-3}$ at 90\% confidence level, which lends support to the hypothesis that the $f_0(1710)$ has a large glueball component.

\end{abstract}

\maketitle
%%%%%%%%%%%%%%%%%%%%%%%%%%%%
%%% introduction %%%
\section{INTRODUCTION}

Confinement is a unique property of quantum chromodynamics (QCD), and can be probed via the spectrum of mesons.
While the quark model describes a conventional meson as a bound state of 
    a quark and an antiquark,
lattice QCD (LQCD) and QCD-motivated models predict a more rich spectrum of mesons that 
includes bound states with gluonic degrees of freedom, 
such as glueballs and hybrids.
Radiative decays of the $\jpsi$ meson provide a gluon-rich environment and are therefore regarded as one of the most promising hunting grounds for gluonic excitations~\cite{Cakir:1994jf,Close:1996yc,Sarantsev:2021ein,Rodas:2021tyb}. 

The spectrum of glueballs, states composed of only gluons, is predicted by quenched LQCD~\cite{b1,b2,b3}.  
The lightest glueball is expected to have scalar quantum numbers, $J^{PC} = 0^{++}$, and a mass between 1.5 and 1.7~GeV/$c^{2}$~\cite{b2,b3}.
LQCD calculations also predict that the branching fraction of the $\jpsi$ radiatively decaying into a pure scalar glueball is 3.8(9)$\times$10$^{-3}$ ~\cite{glueballBR,Sarantsev:2021ein,Rodas:2021tyb}. 
The $f_0(1710)$ is a strong candidate for the lightest glueball, but this identification is complicated by possible mixing with conventional mesons.
Recent partial wave analyses (PWA) of  $\jpsi \rightarrow \gamma\eta\eta$~\cite{getaeta} and $\jpsi \rightarrow \gamma$$K_{S} K_{S}$~\cite{gkk} by BESIII show that the
product branching fractions $\BR(\jpsi \rightarrow \gamma f_0)$ $\times$$\BR(f_0 \rightarrow \eta\eta $  or $ K_{S} K_{S})$
 are one order of magnitude larger for the $f_{0}(1710)$ than for the $f_{0}(1500)$.
Furthermore, the total measured branching fraction of $\jpsi \rightarrow \gamma f_{0}(1710)$, summing over all observed $f_0(1710)$ decay modes, is currently 1.7$\times$10$^{-3}$~\cite{1710BR}, which is compatible with LQCD calculations for a scalar glueball~\cite{glueballBR}. 
Since glueball decays to the $\eta\eta'$ final state are suppressed due to gauge duality~\cite{etaetapBR2}, the $\eta\eta'$ final state is a crucial probe for distinguishing glueballs from conventional mesons~\cite{etaetapBR1}.  
Assuming the glueball branching ratio $\BR(G\rightarrow KK)$/$\BR(G\rightarrow\pi\pi)$ 
is within the range of those measured for the $f_{0}(1710)$ in the Review of Particle Physics by
Particle Data Group (PDG)~\cite{Zyla:2020zbs}, 
Ref.~\cite{etaetapBR2}
predicts the ratio $\BR(G\rightarrow \eta\etap)$/$\BR(G\rightarrow\pi\pi)$ to be less than 0.04.

Hybrid mesons are an additional type of QCD exotic state with gluonic degrees of freedom. 
They were first proposed several decades 
ago~\cite{Horn:1977rq,Isgur:1984bm,Chanowitz:1982qj,Barnes:1982tx,Close:1994hc}, 
and have been the source of more recent
LQCD~\cite{Lacock:1996ny,MILC:1997usn,Dudek:2011bn,Dudek:2013yja} and phenomenological QCD studies~\cite{Szczepaniak:2001rg,Szczepaniak:2006nx,Guo:2008yz,Bass:2018uon}. 
Models and LQCD predict that the exotic $J^{PC} = 1^{-+}$ nonet of hybrid mesons 
is the lightest, 
with a mass around 1.7 -- 2.1~GeV/$c^{2}$~\cite{Meyer:2015eta,Dudek:2013yja,Lacock:1996ny}. 
The predicted decay widths are model-dependent; most hybrids are expected to be rather broad, but some can be as narrow as 100 MeV~\cite{Page:1998gz}. 
To date, there has been experimental evidence for as many as three isovector states with exotic quantum numbers $J^{PC} = 1^{-+}$:
the $\pi_{1}(1400)$, $\pi_{1}(1600)$, and $\pi_{1}(2015)$~\cite{Meyer:2010ku, Klempt:2007cp,JPAC:2018zyd,Woss:2020ayi}.
Finding an isoscalar $1^{-+}$ hybrid state is critical for establishing the hybrid multiplet. Decaying to $\eta\eta'$ in a P-wave is expected for an isoscalar $1^{-+}$ hybrid state~\cite{bib_etaetap_Pwave_1,bib_etaetap_Pwave_2,Eshraim:2020ucw}.
%Finding an isoscalar $1^{-+}$ hybrid state, which can decay to $\eta\eta'$ in a P-wave~\cite{bib_etaetap_Pwave_1,bib_etaetap_Pwave_2},is critical for establishing the hybrid nonet.
%Tetraquark and molecular states may also have $J^{PC} = 1^{-+}$~\cite{Meyer:2010ku}.
% It is important to search for isoscalar state with exotic quantum numbers $1^{-+}$ in gluon-rich radiative $\jpsi$ decays.

In this paper, based on a sample of (10.09$\pm$0.04)$\times$10$^{9}$ $\jpsi$ events collected with the BESIII detector~\cite{Ablikim:NumOfJpsi}, we present a partial wave analysis of $J/\psi\rightarrow\gamma\eta\eta'$ to search for $1^{-+}$ and investigate the decay property of $f_0(1710)$.
 The $\eta$ is reconstructed via the decay channel $\gamma\gamma$, and the $\eta'$ is reconstructed via the decay channels  $\eta'\rightarrow\gamma\pimp$ and $\eta'\rightarrow\eta\pimp$. This paper is accompanied by a letter submitted to Physical Review Letters ~\cite{PRL}.

%%%%%%%%%%%%%%%%%%%%%%%%%%%%
%%% BESIII detector and MC Simulation %%%%%%%%%%%%%%

\section{BESIII DETECTOR AND MONTE CARLO SIMULATION}

%%% BESIII detector
The BESIII detector~\cite{Ablikim:2009aa} records symmetric $e^+e^-$ collisions provided by the BEPCII storage ring, which operates with a peak luminosity of $1\times10^{33}$~cm$^{-2}$s$^{-1}$  at the center-of-mass energy 3.89~GeV. BESIII has collected large data samples in           the energy region between 2.0 and 4.9~GeV~\cite{Ablikim:2019hff}. The cylindrical core of the BESIII detector covers 93\% of the full solid angle and consists of a helium-based multilayer drift chamber~(MDC), a plastic scintillator time-of-flight system~(TOF), and a CsI(Tl) electromagnetic calorimeter~(EMC), which are all enclosed in a superconducting solenoidal magnet providing a 1.0~T (0.9~T in 2012) magnetic field. The solenoid is supported by an octagonal flux-return yoke with resistive plate counter muon identification modules interleaved with steel. 
The charged-particle momentum resolution at $1~{\rm GeV}/c$ is $0.5\%$, and the $dE/dx$ resolution is $6\%$ for electrons The EMC measures photon energies with a resolution of $2.5\%$ ($5\%$) at $1$~GeV in the barrel (end cap) region. The time resolution in the TOF barrel region is 68~ps, while that in the end cap region is 110~ps. The end cap TOF system was upgraded in 2015 using multigap resistive plate chamber technology, providing a time resolution of 60~ps~\cite{etof1,etof2,etof3}.

%%% simulation
Simulated data samples produced with a {\sc geant4}-based~\cite{geant4} Monte Carlo (MC) package, which includes the geometric description of the BESIII detector and the detector response, are used to optimize the event selection criteria, to determine detection efficiencies, and to estimate backgrounds.  Signal MC samples for the process $\jpsi\rightarrow\gamma\eta\etap$ with the subsequent decays $\eta\rightarrow\gamma\gamma$ and $\etap\rightarrow\eta\pimp$ are generated uniformly in phase space. The decay $\etap\rightarrow\gamma\pimp$ is simulated by taking into account both $\rho-\omega$ interference and the box anomaly~\cite{bibDIY}.

An inclusive  MC sample with 10.01$\times$10$^{9}$ $\jpsi$  decays is used to study backgrounds. 
The known decay modes are modeled with EVTGEN~\cite{bib17}  by incorporating branching fractions taken from the Particle Data Group~\cite{Zyla:2020zbs}, and the remaining unknown decays are generated using the LUNDCHARM~\cite{bib18} generator. The simulation includes the beam energy spread and initial state radiation (ISR) in the $e^+e^-$ annihilations modeled with the generator {\sc kkmc}~\cite{ref:kkmc}. Final state radiation (FSR) from charged particles is incorporated with the PHOTOS package~\cite{photos}.

%%% Event Selection %%%%%%%%
\section{EVENT SELECTION}

%%% Event Selection : charged track %%%%%%%%
Charged tracks are reconstructed from hits in the MDC and are required to have $|$cos$\theta | \textless$ 0.93, where $\theta$ is the polar angle defined with respect to the symmetry axis of the MDC in the laboratory frame. Tracks must approach within 10 cm of the interaction point in the beam direction and 1 cm in the plane perpendicular to the beam where the distances are defined in the laboratory frame. Each track is assumed to be a pion, and no particle identification is applied.

%%% Event Selection : photon %%%%%%%%

Photon candidates are required to have 
 energy deposition above 25~MeV in the barrel region ($|$cos$\theta |$$\textless$0.80) or 50~MeV in the end caps (0.86$\textless$$|$cos$\theta |$$\textless$0.92). To exclude spurious photons caused by hadronic interactions and final state radiation, photon candidates must be at least 10$^{o}$ away from any charged tracks when extrapolated to the EMC. To suppress spurious photons due to electronic noise or energy deposits unrelated to the event, candidate showers are required to occur within 700~ns of the event start time.

\begin{figure*}[htbp]
  \centering
  \subfigure{
  \label{gpp select etap}
  	\includegraphics[width=0.4\textwidth,height=0.2\textheight]{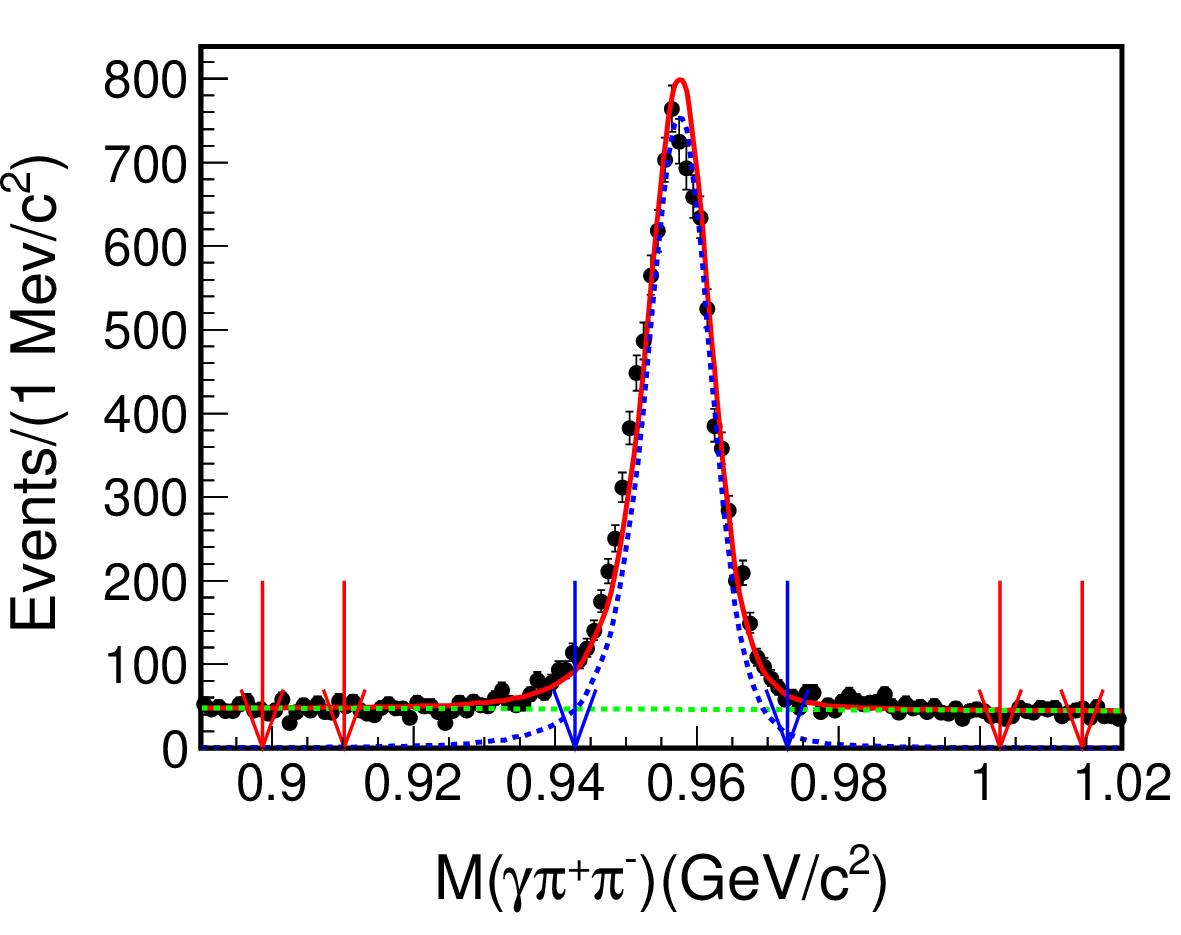}\put(-45,100){(a)}}%
  \subfigure{
  \label{epp select etap}
    \includegraphics[width=0.4\textwidth,height=0.2\textheight]{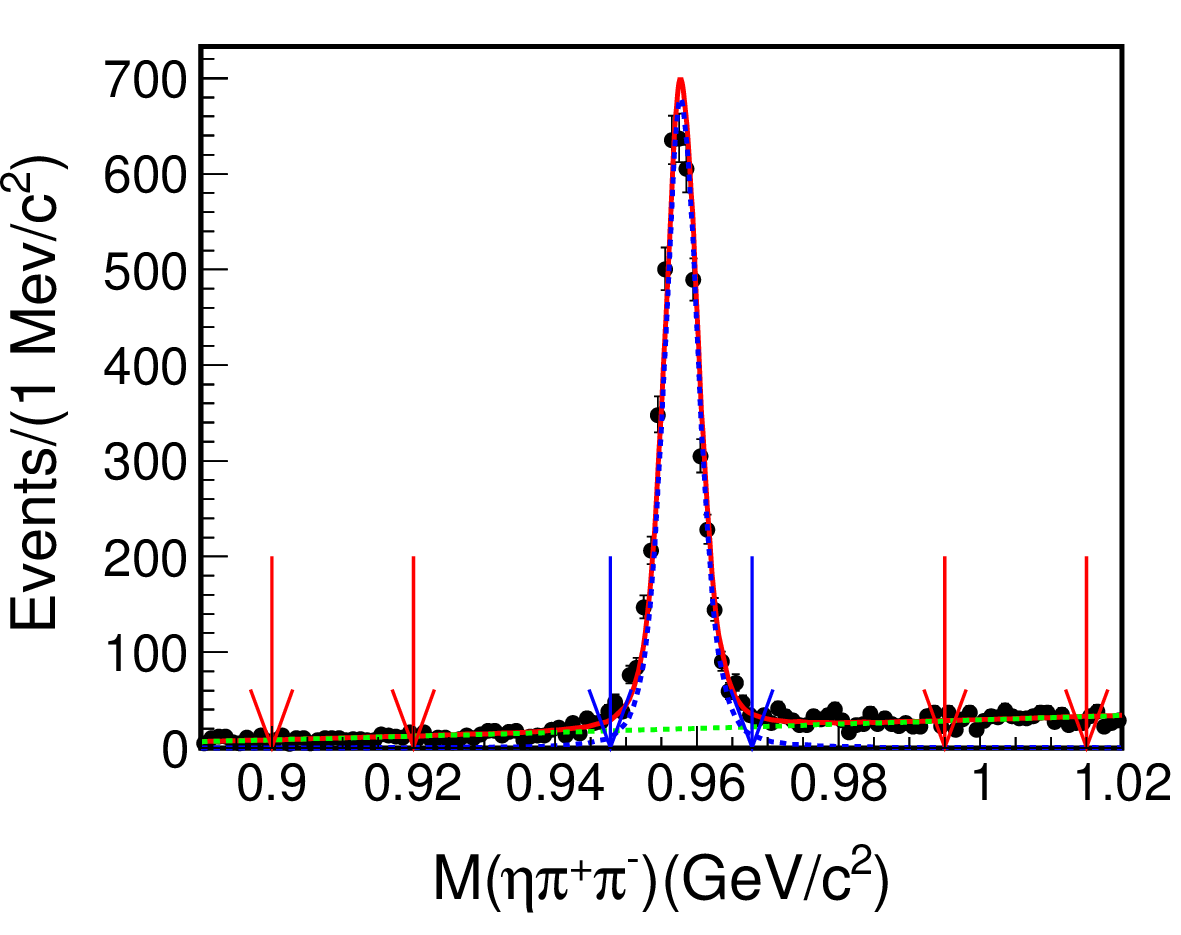}\put(-45,100){(b)}} %
  \subfigure{
  \label{gpp etaetap}
    \includegraphics[width=0.4\textwidth,height=0.2\textheight]{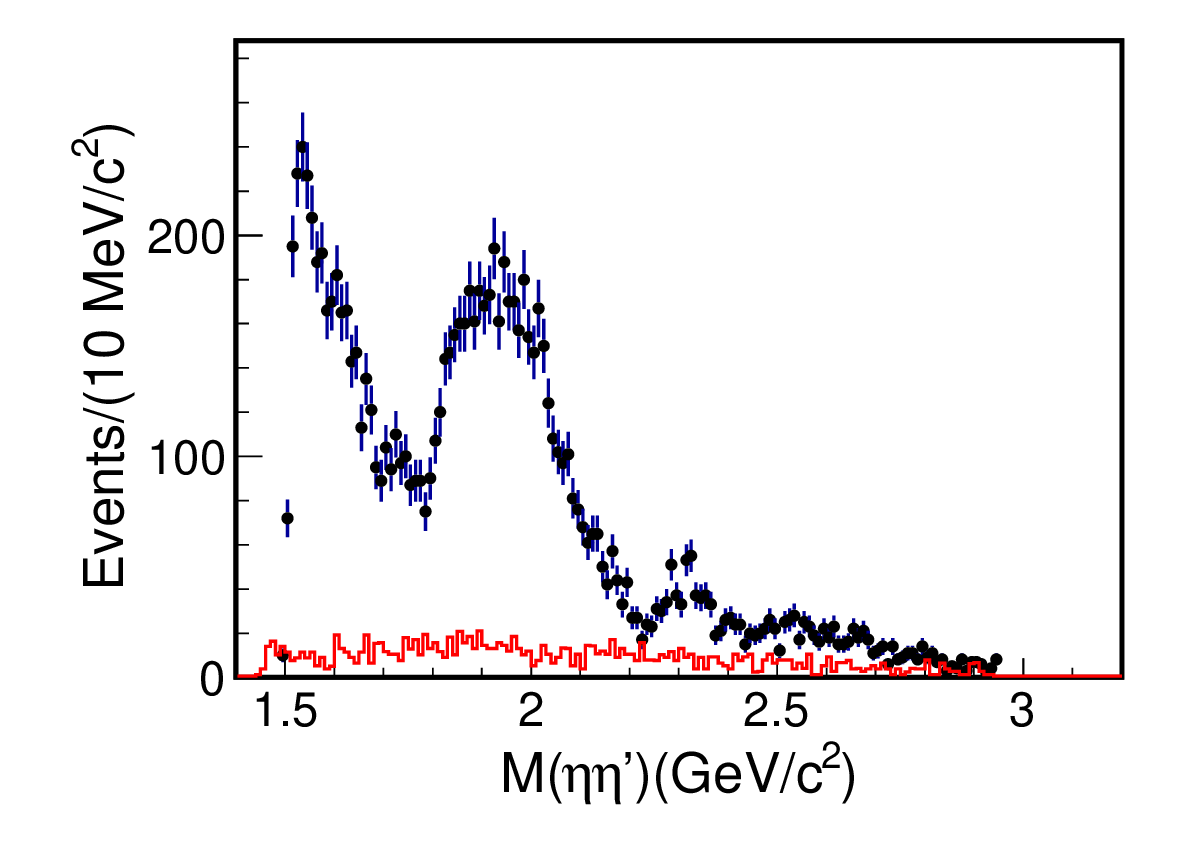}\put(-45,100){(c)}}%
  \subfigure{
  \label{epp etaetap}
    \includegraphics[width=0.4\textwidth,height=0.2\textheight]{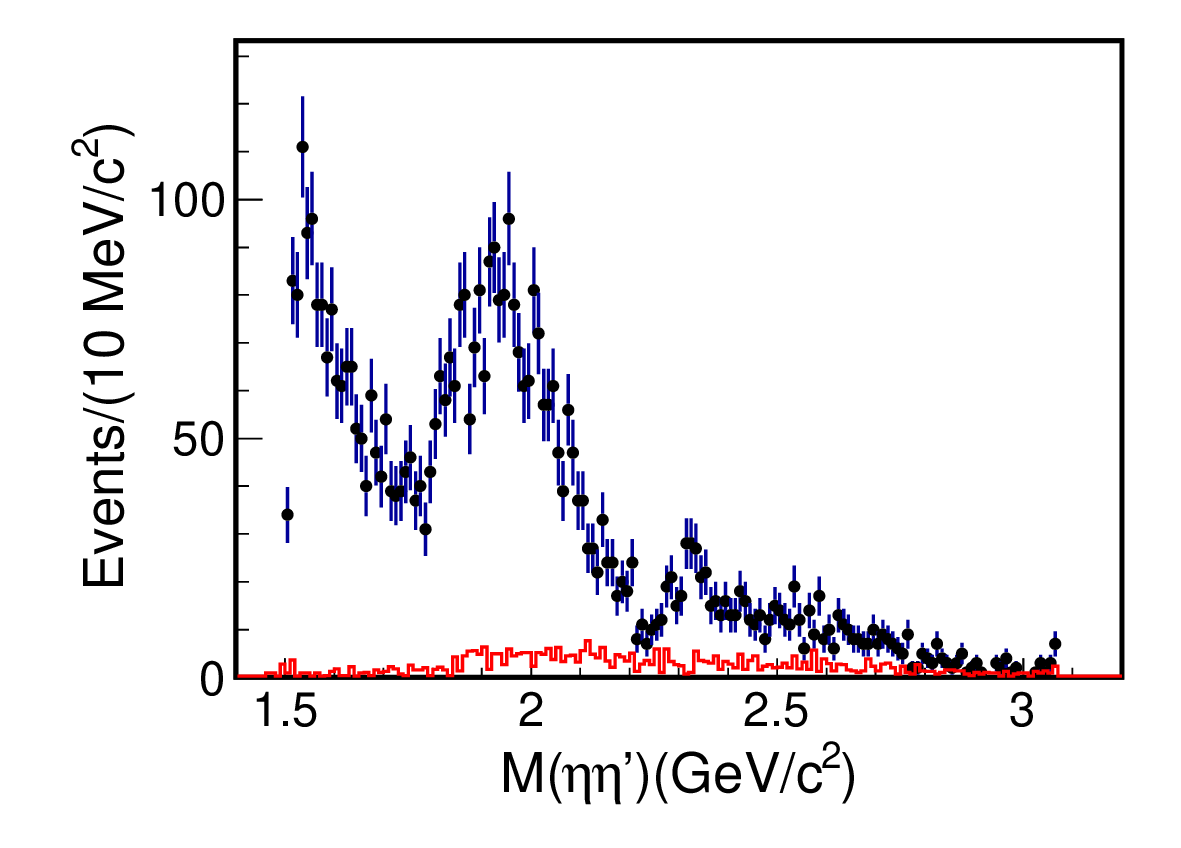}\put(-45,100){(d)}} %
  \subfigure{
  \label{gpp dalitz}
    \includegraphics[width=0.4\textwidth,height=0.2\textheight]{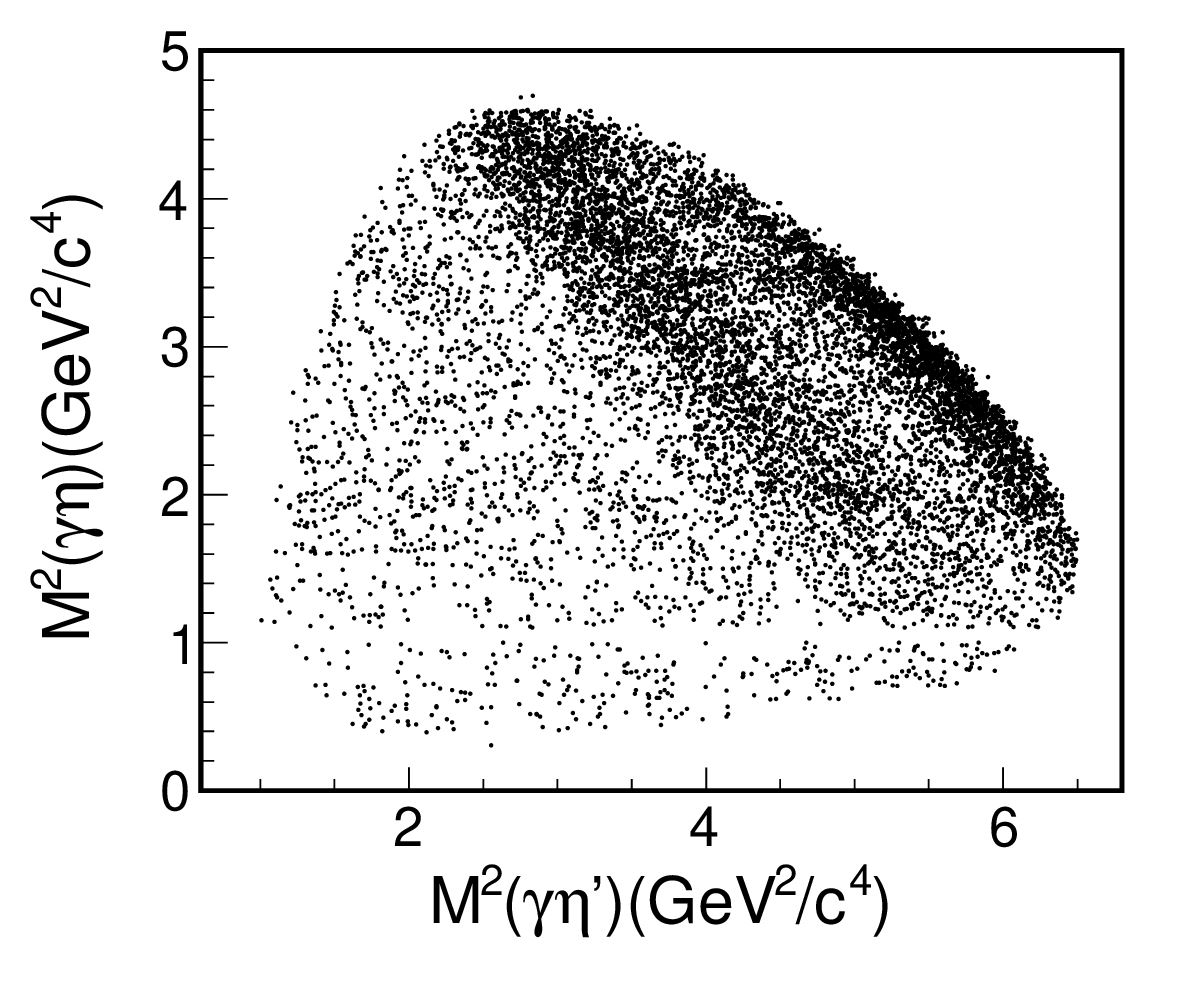}\put(-45,100){(e)}}%
  \subfigure{
  \label{epp etaetap}
     \includegraphics[width=0.4\textwidth,height=0.2\textheight]{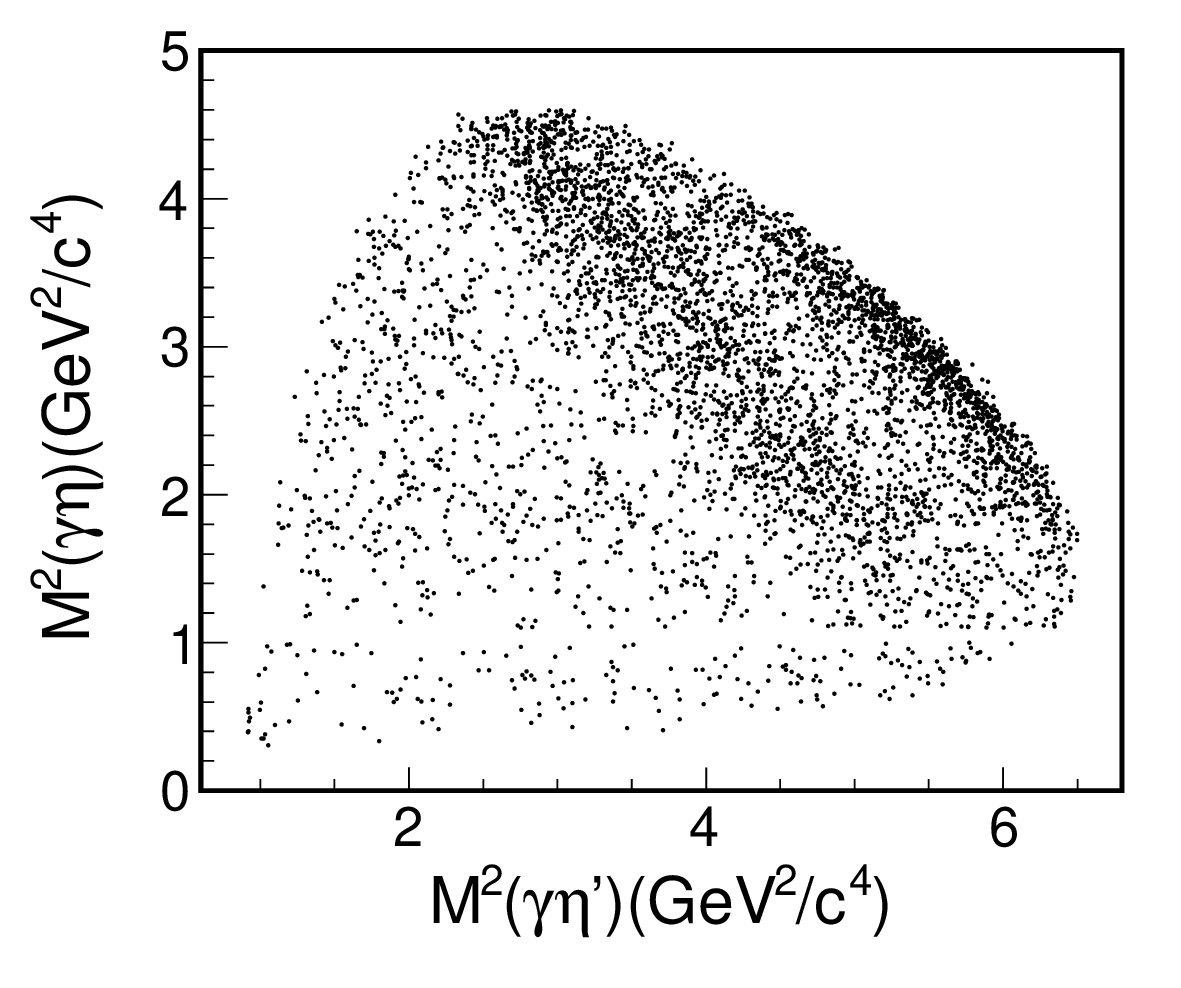}\put(-45,100){(f)}}\\%

   \caption{
Invariant mass distributions for 
(a,c,e)~$\jpsi\rightarrow\gamma\eta\eta', \eta'\rightarrow\gamma\pimp$ and 
(b,d,f)~$\jpsi\rightarrow\gamma\eta\eta', \eta'\rightarrow\eta\pimp$. 
(a,b)~Invariant mass distributions for $\eta'$ candidates. 
The events between the blue arrows are the selected signal events, and 
events between the red arrows on the same side are selected sideband events. 
The black points with error bars are data, 
the red solid line is the fit result, 
the blue dashed line is the $\etap$ signal shape from signal MC, 
and the  green dashed line is the 2nd degree polynomial function for the background. 
(c,d)~Invariant mass distributions of the $\eta\eta'$ for the selected $\gamma\eta\eta'$ candidates. 
The points with error bars are data and the red line shows the background events estimated from the $\eta'$ sideband. 
(e,f)~The corresponding Dalitz plots for the selected $\gamma\eta\eta'$ candidates. }
  \label{final etaetap and Dalitz}
\end{figure*}

%%% Event Selection : gamma pi pi %%%%%%%%
For the $\jpsi\rightarrow\gamma\eta\etap$, $\etap\rightarrow\gamma\pimp$ channel, events are reconstructed with two oppositely charged tracks and at least four candidate photons. A five-constraint (5C) kinematic fit under the hypothesis $\jpsi\rightarrow\gamma\gamma\eta\pimp$ is performed by constraining energy-momentum conservation and the mass of one pair of photons to the nominal mass of the $\eta$ ($m_{\eta}$) from the PDG~\cite{Zyla:2020zbs}. If there is more than one combination, the combination with the minimum $\chi^{2}_{\rm 5C}$ is retained. The resulting $\chi^{2}_{\rm 5C}$ is required to be less than 15. 
To suppress backgrounds from processes with three or five photons in the final state, four-constraint (4C) kinematic fits are performed by constraining energy-momentum conservation under the hypotheses $\jpsi\rightarrow 3\gamma\pimp$, $\jpsi\rightarrow 4\gamma\pimp$, and $\jpsi\rightarrow 5\gamma\pimp$. 
The $\chi^{2}_{\rm 4C}$(4$\gamma\pip\pim)$ is required to be less than all possible $\chi^{2}_{\rm 4C}$(3$\gamma\pip\pim$) and $\chi^{2}_{\rm 4C}$(5$\gamma \pip \pim$). 
To reconstruct the $\etap$ candidate, the $\gamma\pip\pim$ combination with the minimum $\left|M(\gamma\pip\pim)-m_{\etap}\right|$ is chosen, where $m_{\etap}$ is the nominal mass of the $\etap$ taken from the PDG~\cite{Zyla:2020zbs}. 
The invariant mass distribution of $\gamma\pip\pim$ is shown in Fig.~\ref{gpp select etap}. Events with  $\left|M(\gamma\pip\pim)-m_{\etap}\right|\textless$ 0.015~GeV/$c^{2}$ are selected for further analysis. The $\pimp$ invariant mass is required to be near the $\rho^{0}$ mass region, 0.6 $\textless M(\pip\pim)\textless$ 0.8~GeV/$c^{2}$. 
To suppress backgrounds containing  $\pio$ and backgrounds due to misreconstructed $\eta$, 
events with $|M(\gamma\gamma)-m_{\pio}|\textless$ 0.04 or $\left|M(\gamma\gamma)-m_{\eta}\right|\textless$ 0.02~GeV/$c^{2}$ are rejected, where $M(\gamma\gamma)$ are the invariant masses of all photon pairs except the photon pair assigned to the $\eta$ and $m_{\pio}$ is the nominal mass of the $\pio$~\cite{Zyla:2020zbs}. 
There is a clear $\phi$ signal in the $\gamma\eta$ invariant mass distribution corresponding to $\jpsi\rightarrow\phi\etap, \phi\rightarrow\gamma\eta$. Since detector resolution is difficult to model in partial wave analyses, and since the process $\jpsi\rightarrow\phi\etap$ is not our primary interest, we exclude it by rejecting events with $|M(\gamma\eta)-m_{\phi}|\textless$ 0.04~GeV/$c^{2}$. According to the study of inclusive MC,  in order to further reduce the background from $J/\psi\rightarrow3\gamma\pi^{+}\pi^{-}$ with spurious photons, only events with $M(\eta\etap)\textless$2.95~GeV/$c^{2}$ are chosen for further analysis.

%%% Event Selection : eta pi pi %%%%%%%%

For the $\jpsi\rightarrow\gamma\eta\etap$, $\etap\rightarrow\eta\pimp$ channel, 
events are reconstructed with two oppositely charged tracks and at least five candidate photons. 
A six-constraint (6C) kinematic fit under the hypothesis $\jpsi\rightarrow\gamma\eta\eta\pimp$ is performed by constraining energy-momentum conservation and the masses of two pairs of photons to $m_{\eta}$. 
If there is more than one combination, the combination with the minimum $\chi^{2}_{\rm 6C}$ is retained. The resulting $\chi^{2}_{\rm 6C}$ is required to be less than 45. 
To suppress backgrounds with four or six photons in the final state, 4C kinematic fits are performed by constraining energy-momentum conservation under the hypotheses $\jpsi$$\rightarrow$$4\gamma\pimp$, $\jpsi$$\rightarrow$$5\gamma\pimp$, and $\jpsi$$\rightarrow$$6\gamma\pimp$. 
The $\chi^{2}_{\rm 4C}$($5\gamma\pip\pim)$ is required to be less than all possible $\chi^{2}_{\rm 4C}$($4\gamma\pip\pim$) and  $\chi^{2}_{\rm 4C}$($6\gamma\pip\pim$). 
The $\eta\pip\pim$ combination with the minimum $\left|M(\eta\pip\pim)-m_{\etap}\right|$ is used to reconstruct the $\etap$ candidate. The invariant mass distribution of $\eta\pip\pim$ is shown in Fig.~\ref{epp select etap}. Events with  $\left|M(\eta\pip\pim)-m_{\etap}\right|\textless$ 0.01~GeV/$c^{2}$ are selected for further analysis. 
Backgrounds containing a $\pio$ and backgrounds due to misreconstructed $\eta$ are suppressed by rejecting events with $|M(\gamma\gamma)-m_{\pio}|\textless$ 0.03 or $\left|M(\gamma\gamma)-m_{\eta}\right|\textless$ 0.02~GeV/$c^{2}$, respectively. 
To exclude $\jpsi\rightarrow\phi\etap, \phi\rightarrow\gamma\eta$, events with $|M(\gamma\eta)-m_{\phi}|\textless$ 0.04~GeV/$c^{2}$ are rejected.

The invariant mass distributions of $\eta\eta'$ and the Dalitz plot for the selected $\gamma\eta\eta'$ candidate events from the two decay channels are shown in Fig.~\ref{final etaetap and Dalitz}(c, d) and Fig.~\ref{final etaetap and Dalitz}(e, f), respectively. Clear structures in the $\eta\eta'$ invariant mass spectrum are observed.%%% background analysis

Potential backgrounds are studied using the inclusive MC sample of 10.01$\times$10$^{9}$ $\jpsi$ decays (as described in Sec. II). 
No significant peaking background is observed in the invariant mass distribution of the $\eta'$. 
Backgrounds are estimated by the $\eta'$ sidebands in the data. 
The sideband regions of $\etap \rightarrow \eta \pip \pim$ and $\etap \rightarrow \gamma \pip \pim$ are defined as 
$M(\eta\pip\pim)$ $\in \lbrack 0.900,0.920\rbrack \bigcup \lbrack 0.995,1.015 \rbrack $~GeV/$c^{2}$ and
$M(\gamma\pip\pim)$ $\in \lbrack 0.903,0.915 \rbrack \bigcup \lbrack 1.000,1.012 \rbrack$~GeV/$c^{2}$, respectively. 
The normalization factors for events in the two sideband regions are obtained by a fit to data of the invariant mass spectrum of $M(\eta\pip\pim)$ and $M(\gamma\pip\pim)$. The $\etap$ signal shapes are determined from the shapes of signal MC samples described with RooHistPdf~\cite{Verkerke:2003ir}, 
and the backgrounds are described by 2nd degree polynomial functions. The definition of the sidebands and the fit results are shown in Fig.~\ref{final etaetap and Dalitz}(a) and Fig.~\ref{final etaetap and Dalitz}(b). 
The background levels for $\jpsi \rightarrow \gamma\eta\etap, \etap \rightarrow \eta \pip \pim$ and $\jpsi \rightarrow \gamma\eta\etap, \etap \rightarrow \gamma \pip \pim$ estimated by the $\etap$ sidebands are 8.3$\%$ and 13.1$\%$, respectively.

%%% PWA %%%%%
\section{PARTIAL WAVE ANALYSIS}
%%% PWA method %%%%%%%%%%
After event selection, the numbers of remaining events for $\jpsi \rightarrow \gamma\eta\etap, \etap \rightarrow \eta \pip \pim$ and $\jpsi \rightarrow \gamma\eta\etap, \etap \rightarrow \gamma \pip \pim$ are 4788 and 10\,544, respectively. The
four-momenta of the reconstructed $\gamma$, $\eta$, and $\etap$
are used to perform the PWA fit.

\subsection{Analysis method}
Using the GPUPWA framework~\cite{gpuframework}, a combined PWA fit is performed to the selected samples of $\jpsi$ $\rightarrow$ $\gamma\eta\etap$, $\etap$ $\rightarrow$ $\gamma\pimp$ and $\jpsi$ $\rightarrow$ $\gamma\eta\etap$, $\etap$ $\rightarrow$ $\eta\pimp$. 
Quasi two-body amplitudes in the sequential radiative decay processes $\jpsi \rightarrow \gamma X, X\rightarrow\eta\etap$ and hadronic decay process $\jpsi \rightarrow\eta X, X\rightarrow \gamma\etap$ and $\jpsi \rightarrow\etap X, X\rightarrow \gamma\eta$ are constructed using the covariant tensor amplitudes described in Ref.~\cite{bib19}. 
Let A$_X$ be the amplitude for a $\jpsi$ decay process including intermediate resonance $X$.
For $\jpsi$ radiative decays, the general form of A$_X$ is

\begin{eqnarray}\label{general amplitudes radiative}
A_X = \psi_{\mu}(m_{1})e_{\nu}^{*}(m_{2})\sum_{k}\Lambda_{k}U_{k}^{\mu\nu},
\end{eqnarray}
where the summation is over the number of independent amplitudes
and for $\jpsi$ hadronic decays, the  general form of A$_X$ is 
\begin{eqnarray}\label{general amplitudes hadronnic}
A_X  = \psi_{\mu}(m_1)\sum_{k}\Lambda_{k}U_{k}^{\mu},
\end{eqnarray}
where $\psi_{\mu}(m_{1})$ is the polarization four-vector for the $\jpsi$; 
$e_{\nu}^{*}(m_{2})$ is the polarization four-vector for the photon; $m_{1}$ and $m_{2}$ are the spin projections of the $\jpsi$ and photon, respectively;
$U_{k}^{\mu\nu}$ is the $k$th independent partial wave amplitude of $\jpsi$ radiative decays to intermediate resonance $X$ with coupling strength determined by a complex parameter  $\Lambda_{k}$; 
and $U_{k}^{\mu}$ is the $k$th independent partial wave amplitude of $\jpsi$ hadronic decays to intermediate resonance $X$ with coupling strength determined by a complex parameter  $\Lambda_{k}$.
The partial wave amplitudes $U_{k}^{\mu\nu}$ and $U_{k}^{\mu}$ are constructed using the four-momenta of the reconstructed $\gamma$, $\eta, \etap$.  %Their specific expressions are given in Ref.~\cite{bib19}.

The amplitudes for the $J/\psi$ radiative decay processes $\jpsi \rightarrow \gamma f_0, \gamma f_2, \gamma f_4$ are given in Ref.~\cite{bib19}. For $\jpsi \rightarrow \gamma \eta_1$, 
where the $\eta_1$ is an isoscalar state with exotic quantum numbers $J^{PC}=1^{-+}$, the $\eta_1$ can decay into $\eta\etap$ in a P-wave~\cite{bib_etaetap_Pwave_1,bib_etaetap_Pwave_2} with two amplitudes:
\begin{eqnarray}\label{gamma eta1 -- 1}
U_{(\gamma\eta_{1})1}^{\mu\nu} = g^{\mu\nu}p^{\alpha}_{\psi}\tilde{t}_{\alpha}^{(\eta_{1})}f^{(\eta_{1})},
\end{eqnarray}
\begin{eqnarray}\label{gamma eta1 -- 2}
U_{(\gamma\eta_{1})2}^{\mu\nu} = q^{\mu}\tilde{t}^{(\eta_{1})\nu}f^{(\eta_{1})},
\end{eqnarray}
where $g^{\mu\nu}$ is the metric tensor, $p_{\psi}$ is the four-momentum of the $\jpsi$, $q$ is the four-momentum of the radiative photon, and $f^{(\eta_{1})}$ is the  propagator for the process  $\eta_1\rightarrow\eta\etap$. Blatt-Weisskopf barrier factors~\cite{Chung:1993da1,Chung:1993da2,VonHippel:1972fg} are included in the orbital angular momentum covariant tensors $\tilde{t}$. Due to the special properties (massless and gauge invariance) of the photon, the number of independent partial wave amplitudes for a  $J/\psi$ radiative decay is smaller than for the corresponding decay to a massive vector meson, the details are given in Ref.~\cite{bib19}.

For the $J/\psi$ hadronic decay processes  $\jpsi \rightarrow \rm V \eta', V\rightarrow\gamma\eta$, where V is vector meson that has quantum numbers $J^{PC}=1^{--}$, such as $\phi$, $\rho$, $\omega$ and their excitations, the corresponding amplitude is
\begin{eqnarray}\label{phi etap}
U_{{\rm{(V\etap)}}\rm P}^{\mu\nu} =
 g^{\mu\mu'}\epsilon_{\mu'\nu'\alpha\beta}p_{\psi}^{\alpha}\tilde{T}^{\beta}_{(\rm V\etap)}\epsilon^{\nu'\lambda\sigma\nu}q_{\lambda}p_{1\sigma}f^{(\rm V)},
\end{eqnarray}
where, $\epsilon_{\mu'\nu'\alpha\beta}$ is the totally antisymmetric tensor, $p_{1}$ is the four-momentum of the $\eta'$  from the $J/\psi$ decay, $\tilde{T}^{\beta}_{\rm V\etap}$ is the orbital angular momentum covariant tensor of the process $\jpsi$ $\rightarrow$ $\rm V \eta^{\prime}$, and
$f_{\gamma\eta}^{(\rm V)}$ is the propagator for the process $\rm V \rightarrow \gamma \eta$.
The subscript P indicates $J/\psi\rightarrow {\rm{V}}\eta^{\prime}$ is in a P wave.
The amplitude of  $\jpsi \rightarrow \rm V \eta, V\rightarrow\gamma\eta^{\prime}$ is analogously to Eq.~\ref	{phi etap}.
For the process $\jpsi$ $\rightarrow$ $h_{1} \eta'$, $h_{1}$ $\rightarrow$ $\gamma \eta$, the corresponding two independent amplitudes are
\begin{eqnarray}\label{h1 etap -- 1}
U_{(h_{1}\etap)\rm S}^{\mu\nu} = \tilde{g}^{\mu\nu}_{(h_{1})}f^{(h_{1})},
\end{eqnarray}

\begin{eqnarray}\label{h1 etap -- 2}
U_{(h_{1}\etap)\rm D}^{\mu\nu} = \tilde{T}^{(2)\mu\alpha}_{(h_{1}\etap)}\tilde{g}^{\nu}_{(h_{1})\alpha}f^{(h_{1})}.
\end{eqnarray}
where $\tilde{g}^{\mu\nu}_{(h_{1})}=g^{\mu\nu}-\frac{p^{\mu}_{h_1}p^{\nu}_{h_1}}{p^{2}_{h_1}}, p_{h_1}$ is the four-momentum of the $h_1$, and $f^{(h_1)}$ is the  propagator for the process  $h_1\rightarrow\gamma\eta$. The superscript (2) on $\tilde{T}$ indicates the orbital momentum between $h_1$ and $\etap$ is 2. The subscripts S and D indicate $J/\psi\rightarrow h_1\eta^{\prime}$ is in an S wave and D wave, respectively. The amplitudes for the process $\jpsi$ $\rightarrow$ $h_{1} \eta$, $h_{1}$ $\rightarrow$ $\gamma \eta^{\prime}$ are analogously
 to Eq.~\ref{h1 etap -- 1} and Eq.~\ref{h1 etap -- 2}.

In this analysis, resonance decays are described by a Breit-Wigner (BW) function, parametrized by a constant-width, relativistic BW propagator,
\begin{eqnarray}\label{general amplitudes}
f^{\rm (X)} = \frac{1}{M^2 - s - \frac{1}{c^{2}}iM\Gamma},
\end{eqnarray}
where $M$ and $\Gamma$ are the mass and width of the intermediate resonance X, and $\sqrt{s}$ is the invariant mass of the $\eta\etap$, $\gamma\eta$, or $\gamma\etap$ system.

The complex coefficients of the amplitudes (relative magnitudes and phases) and resonance parameters (masses and widths) are determined by an unbinned maximum likelihood fit to the data.  The likelihood is constructed following a method similar to that used in Ref.~\cite{Ablikim:2016hlu}.

The probability to observe the $i$th event characterized by the measurement $\xi_{i}$, i.e., the measured four-momenta of the particles in the final state, is

\begin{eqnarray}\label{Prob ith event}
P(\xi_{i}) = \frac{\left|M(\xi_{i}) \right|^2\epsilon(\xi_{i})\Phi(\xi_{i})}{\sigma^\prime},
\end{eqnarray}
where $\epsilon(\xi_{i})$ is the detection efficiency, $\Phi(\xi_{i})$ is the standard element of phase space, and $M(\xi_{i})= \sum_{X} A_X(\xi_i)$ is the matrix element describing the decay processes from the $J/\psi$ to the final state $\gamma\eta\eta^\prime$.  $A_X(\xi_i)$ is the amplitude corresponding to intermediate resonance X as defined in Eq.~\ref{general amplitudes radiative} and Eq~\ref{general amplitudes hadronnic}.
$\sigma^\prime\equiv \int{\left|M(\xi) \right|^2\epsilon(\xi)\Phi(\xi)d\xi}$
is the normalization integral.

The joint probability  for observing $N$ events in the data sample is
\begin{small}
\begin{eqnarray}\label{joint probability density}
\mathcal{L}  = \prod\limits_{i=1}^{N}\frac{ \left|M(\xi_{i}) \right|^2\epsilon(\xi_{i})\Phi(\xi_{i})  }{\sigma^\prime}.
\end{eqnarray}
\end{small}

For technical reasons, rather than maximizing $\mathcal{L}$, $-{\rm ln}\mathcal{L}$ is minimized, with

\begin{small}
\begin{eqnarray}\label{minus ln L}
 -{\rm ln}\mathcal{L} = -\sum\limits_{i=1}^{N}{\rm ln}  \left|M(\xi_{i}) \right|^2 + N{\rm ln}\sigma^\prime - \sum\limits_{i=1}^{N}  {\rm ln}\epsilon(\xi_{i})\Phi(\xi_{i}),
\end{eqnarray}
\end{small}
for a given dataset. The third term is a constant and has no impact on the determination of the parameters of the amplitudes or on the relative changes of $-{\rm ln}\mathcal{L}$ values. In the fitting, the third term will not be considered.

The free parameters are optimized by MINUIT~\cite{minuit}. The normalization integral $\sigma'$ is evaluated using MC with importance sampling~\cite{MCSampling2,MCSampling1}. An MC sample of $N_{\rm gen}$ is generated with signal events distributed uniformly in phase space. These events are put through the detector simulation, subjected to the selection criteria and yield a sample of $N_{\rm acc}$ accepted events. The normalization integral is computed as:

\begin{small}
\begin{eqnarray}\label{MCIntegral}
\sigma' = \int{ \left|M(\xi) \right|^2\epsilon(\xi)\Phi(\xi)d\xi}  \propto \frac{1}{N_{\rm gen}}\sum\limits_{j}^{N_{\rm acc}} \left|M(\xi_{j}) \right|^2,
\end{eqnarray}
\end{small}
where the constant value of the phase space integral $\int{\Phi(\xi)d\xi}$ is ignored.

Instead of modeling the background, the likelihood is defined by the signal PDF [Eq.~\ref{minus ln L}] and the contribution to the negative log-likelihood from background events in the signal region is removed by subtracting out the negative log-likelihood of events in the $\etap$ sideband region in proper proportion~\cite{Langenbruch:2019nwe}, i.e.,
\begin{eqnarray}
-{\rm ln}\mathcal{L}_{\rm signal} = - ({\rm ln}\mathcal{L}_{\rm data} - \sum\limits_{i} w_i \cdot {\rm ln}\mathcal{L}^i_{\rm background}),
\end{eqnarray}
where $-{\rm ln}\mathcal{L}_{\rm signal}$ is the likelihood for the signal, ${\rm ln}\mathcal{L}_{\rm data}$ is the likelihood calculated by Eq.~\ref{minus ln L} using the data sample,  ln$\mathcal{L}^i_{\rm background}$ is the likelihood calculated by Eq.~\ref{minus ln L} using the events of $i$th sideband, and $w_i$ is the normalization factor for background events in the $i$th sideband region, which is determined from the fit results of Fig.~\ref{final etaetap and Dalitz}(a) and Fig.~\ref{final etaetap and Dalitz}(b).

The number of fitted events $N_{X}$ for an intermediate resonance $X$ is defined as:
\begin{eqnarray}\label{13}
N_X = \frac{\sigma_X}{\sigma'}\cdot N,
\end{eqnarray}
where $N$ is the number of selected events after background subtraction, and
\begin{eqnarray}\label{14}
\sigma_X = \frac{1}{N_{\rm gen}}\sum_{j}^{N_{\rm acc}}|A_X(\xi_j)|^2,
\end{eqnarray}
is calculated with the same MC sample as the normalization integral $\sigma'$.

\iffalse
For two intermediate resonances $X_1$ and $X_2$, which have $N_{W_1}$ and $N_{W_2}$ independent partial wave amplitudes, $A_{i_1}$ and $A_{i_2}$, respectively, the number of fitted events arising from their interference, denoted as $N_{X_1,X_2}$, is:
\begin{eqnarray}\label{15}
N_{X_1,X_2} = \frac{\sigma_{X_1,X_2}}{\sigma}\cdot N,
\end{eqnarray}
where
\begin{eqnarray}\label{16}
\sigma_{X_1,X_2} = \frac{1}{N_{\rm acc}}\sum_{k}^{N_{\rm acc}}\sum_{j_1}^{N_{W_1}}\sum_{j_1}^{N_{W_2}}((A_{j_1})_k(A^{*}_{j_2})_k + (A^{*}_{j_1})_k(A_{j_2})_k),
\end{eqnarray}
\fi

\iffalse
The branching ratio of $J/\psi\rightarrow \gamma X, X \rightarrow \eta\etap, \etap \rightarrow \eta \pip \pim$ is calculated as:
\begin{eqnarray}\label{17}
\BR(J/\psi\rightarrow \gamma X\rightarrow \gamma \eta\etap) = \frac{N_{X}}{N_{J/\psi}\cdot \epsilon_X \cdot \BR_{\etap\rightarrow \eta\pip\pim}\cdot \BR_{\eta\rightarrow \gamma\gamma}^{2}},
\end{eqnarray}
\fi
The detection efficiency $\epsilon_X$ for an intermediate resonance $X$ is obtained by the partial wave amplitude weighted MC sample,
\begin{eqnarray}\label{19}
\epsilon_X =  \frac{\sum_{j}^{N_{\rm acc}}|A_X(\xi_j)|^2}{\sum_{n}^{N_{\rm gen}}|A_X(\xi_n)|^2},
\end{eqnarray}

A combined unbinned maximum likelihood fit is performed for the two decay channels by adding the negative log-likelihood of signal, $-{\rm ln}\mathcal{L}_{\rm signal}$, for  $\jpsi \rightarrow \gamma\eta\etap$, $\etap\rightarrow\eta\pip\pim$ and that for $\jpsi\rightarrow\gamma\eta\etap$, $\etap\rightarrow\gamma\pip\pim$ together. In the combined fit, the two decay modes share the same set of masses, widths, relative magnitudes, and phases.

The product branching fraction of $\jpsi \rightarrow \gamma X, X\rightarrow \eta\etap$ is obtained according to:
\begin{widetext}
\begin{small}
\begin{eqnarray}
\BR(\jpsi\rightarrow \gamma X) \BR(X\rightarrow  \eta\etap) =  
\frac{N_{X_a}+N_{X_b}}{N_{J/\psi}(\epsilon_{X_a} \cdot \BR(\etap\rightarrow \eta\pip\pim)\BR^2(\eta\rightarrow\gamma\gamma) + \epsilon_{X_b} \cdot \BR(\etap\rightarrow \gamma\pip\pim)\BR(\eta\rightarrow \gamma\gamma))},
\label{combinedBR}
\end{eqnarray}
\end{small}
\end{widetext}
the subscript "$a$" and "$b$" indicate $\jpsi \rightarrow \gamma\eta\etap, \etap \rightarrow \eta \pip \pim$ and $\jpsi \rightarrow \gamma\eta\etap, \etap \rightarrow \gamma \pip \pim$, respectively;  $\epsilon_{X_a}$ and $\epsilon_{X_b}$ are the detection efficiencies of the two different decay channels calculated by Eq.~$\ref{19}$;  $N_{\jpsi}$ is the total number of $\jpsi$ events; and $\BR(\etap\rightarrow \eta\pip\pim)$$=$$(42.5\pm0.5)\%$, $\BR(\eta\rightarrow\gamma\gamma)$=$(39.41\pm0.20)\%$, $\BR$$(\etap$$\rightarrow$$ \gamma\pip\pim)$$=$$(29.5\pm 0.4)\%$
are the branching fractions taken from the PDG~\cite{Zyla:2020zbs}.

The product branching fraction of $\jpsi \rightarrow \eta X, X\rightarrow \gamma\etap$ and $\jpsi \rightarrow \etap X, X\rightarrow \gamma\eta$ can be calculated in a similar way to Eq.~\ref{combinedBR}.

%%%%% PWA results %%%%%%%%%
\subsection{PWA results}

To construct a set of two-body amplitudes to use in the PWA fit,
a ``PDG-optimized'' set of amplitudes is first determined.
To describe the $\eta\etap$ and $\gamma\eta^{(\prime)}$ spectra, 
all kinematically allowed resonances with 
$J^{PC} =$ $0^{++}$, $2^{++}$, and $4^{++}$ (for the $\eta\etap$ system) and
$J^{PC} =$ $1^{+-}$ and $1^{--}$ (for the $\gamma\eta^{(\prime)}$ systems) listed in the PDG~\cite{Zyla:2020zbs} are considered. 
Within the allowed phase space (PHSP) of the $\eta\etap$ system, 
four additional states [the $f_{0}(2102)$, $f_{0}(2330)$, $f_{2}(2240)$, and $f_{4}(2283)$] 
reported in Ref.~\cite{Bugg:2004xu} 
and an additional scalar state (the $f_{0}(1810)$) reported in Ref.~\cite{BESIII:2012rtd} are also considered. 
Table~\ref{Candidate poor} shows the complete set of resonances considered from the PDG, Ref.~\cite{Bugg:2004xu}, and Ref.~\cite{BESIII:2012rtd}. 
All possible sets of amplitudes corresponding to resonances listed in Table~\ref{Candidate poor} are evaluated. 
The statistical significance for each resonance is determined by examining the probability of the change in negative log-likelihood values when this resonance is included or excluded in the fits, 
where the probability is calculated under the $\chi^{2}$ distribution hypothesis taking into account the change in the number of degrees of freedom.
The masses and widths of the resonances near $\eta\eta'$ mass threshold [$f_{0}(1500)$, $f_{2}(1525)$, $f_{2}(1565)$, and $f_{2}(1640)$] 
as well as those with small fit fractions~($\textless$3$\%$)
are always fixed to the PDG~\cite{Zyla:2020zbs} values. 
The mass and width of the $f_{0}(2330)$, which corresponds to a clear structure around 2.3 GeV/$c^{2}$ in the $\eta\eta'$ mass spectrum, are free parameters. All other masses and widths are also free parameters in the fit.
The final PDG-optimized set of amplitudes is the combination where each included resonance has a statistical significance larger than 5$\sigma$. 
Results from the PWA fit using the PDG-optimized set of amplitudes,
including the masses, the widths, and the statistical significances of each component, 
are shown in Table~\ref{PWA solution without eta1, optimize}, 
where the uncertainties are statistical only.

 \begin{table}[!htbp]
\centering
\caption{The set of all intermediate resonances considered when the PDG-optimized set of amplitudes is developed. 
States with quantum numbers $J^{PC} =$ $0^{++}$, $2^{++}$, and $4^{++}$ in the $\eta\etap$ spectrum and states with quantum numbers $J^{PC} =$ $1^{+-}$ and $1^{--}$ in the $\gamma\eta^{(\prime)}$ spectra are considered.}\label{Candidate poor}
\begin{small}
\begin{tabular}{c|cccc}
\hline\hline
 Decay mode  &$0^{++}$           &$2^{++}$      &$4^{++}$  \\
\hline
 &$f_{0}(1500)$                                   & $f_{2}(1525)$      &   $f_{4}(2050)$         \\
 &$f_{0}(1710)$                        & $f_{2}(1565)$  &     $f_{4}(2300)$           \\
  & $f_{0}(1810)$\cite{BESIII:2012rtd}        & $f_{2}(1640)$   &     $f_{4}(2283)$\cite{Bugg:2004xu}       \\
 &  $f_{0}(2020)$                              &  $f_{2}(1810)$   &         \\
 &  $f_{0}(2100)$                                 & $f_{2}(1910)$     &        \\
 $J/\psi\rightarrow \gamma X \rightarrow \gamma \eta\etap$&    $f_{0}(2200)$                                &  $f_{2}(1950)$    &       \\
&    $f_{0}(2330)$                             &   $f_{2}(2010)$   &       \\
&   $f_{0}(2102)$\cite{Bugg:2004xu}                &  $f_{2}(2150)$    &       \\
&   $f_{0}(2330)$\cite{Bugg:2004xu}                  &  $f_{2}(2220)$    &       \\
&                                   &   $f_{2}(2300)$   &       \\
&                                   &  $f_{2}(2340)$    &       \\
&                                   &  $f_{2}(2240)$\cite{Bugg:2004xu}    &       \\

\hline\hline
  &$1^{--}$           &$1^{+-}$      &  \\
\hline
&  $\omega(1420) $                              &  $h_1(1415)$    &            \\
 &  $\omega(1650)$                      & $h_1(1595)$  &               \\
 &     $\phi(1680)$                             &    &          \\
$J/\psi\rightarrow \eta^{(\prime)} X \rightarrow \gamma \eta\etap$ & $\phi(2170)$                               &     &         \\
  & $\rho(1450)$                               &     &         \\
 & $\rho(1700)$                              &     &         \\
 & $\rho(1900)$                               &     &         \\
\hline\hline
\end{tabular}
\end{small}
\end{table}

\begin{table*}[!htbp]
\centering
\linespread{1.5}
\begin{small}
\caption{The mass ($M$), width ($\Gamma$), PDG mass ($M_{\rm PDG}$), PDG width ($\Gamma_{\rm PDG}$), significance (Sig.) and the product branching fractions $\BR(J/\psi\rightarrow\gamma X)\BR(X\rightarrow\eta\etap)$, $\BR(J/\psi\rightarrow \eta^{\prime}X) \BR(X \rightarrow\gamma\eta)$  and
$\BR(J/\psi\rightarrow\eta X)\BR(X\rightarrow\gamma\eta^{\prime})$
 (B.F.) for each component in the PWA fit using the PDG-optimized set of amplitudes. The  uncertainties are statistical.
 }\label{PWA solution without eta1, optimize}
\begin{tabular}{c|ccccccc}
\hline
\hline
Decay mode&Resonance     &$M$ (MeV/$c^{2}$)           &$\Gamma$ (MeV)  &$M$$_{\rm PDG}$ (MeV/$c^{2}$)  &$\Gamma_{\rm PDG}$ (MeV) &B.F. ($\times$10$^{-5}$)    &Sig.                         \\
\hline
&$f_{0}(1500)$   & 1506 & 112     & 1506 & 112               &3.05$\pm$0.07 &$\gg$30$\sigma$                                      \\

&$f_{0}(1810)$   & 1795 & 95      & 1795 & 95                 &0.07$\pm$0.01     &7.6$\sigma$                                      \\

&$f_{0}(2020)$   & 1935$\pm$5  & 266$\pm$9  &1992 &442   &1.67$\pm$0.07 &11.0$\sigma$                                      \\

&$f_{0}(2100)$     & 2109$\pm$11  & 253$\pm$21 &2086 &284   &0.33$\pm$0.03    &5.2$\sigma$                                      \\

 $J/\psi\rightarrow \gamma X \rightarrow \gamma \eta\etap$&$f_{0}(2330)$     & 2327$\pm$4  & 44$\pm$5    &2314 &144    &0.07$\pm$0.01  &8.5$\sigma$                                      \\

&$f_{2}(1565)$    & 1542 & 122   & 1542 & 122              &0.20$\pm$0.03  &6.2$\sigma$                                       \\

&$f_{2}(1810)$    & 1815 & 197   & 1815 & 197              &0.37$\pm$0.03   &7.0$\sigma$                                       \\

&$f_{2}(2010)$    & 2022$\pm$6  & 212$\pm$8  & 2011 & 202 &1.36$\pm$0.10 &8.8$\sigma$                                       \\

&$f_{2}(2340)$    & 2345 & 322   & 2345 & 322               &0.25$\pm$0.04  &6.5$\sigma$                                       \\

&$f_{4}(2050)$    & 2018 & 234   & 2018 & 234               &0.11$\pm$0.02 &5.6$\sigma$                                       \\

\hline
&$h_{1}(1415)$    & 1416 & 90  & 1416 & 90    &0.14$\pm$0.01 &10.3$\sigma$                                       \\

$J/\psi\rightarrow \eta^{\prime} X \rightarrow \gamma \eta\etap$&$h_{1}(1595)$    & 1584 & 384 & 1584 & 384    &0.41$\pm$0.04   &9.7$\sigma$                                       \\

&$\phi(2170)$      & 2160 & 125  & 2160 & 125  &0.24$\pm$0.03  &5.6$\sigma$                                       \\
\hline
$J/\psi\rightarrow \eta X \rightarrow \gamma \eta\etap$&$h_{1}(1595)$   & 1584 & 384 & 1584 & 384    &0.50$\pm$0.03 &11.0$\sigma$                                       \\

&$\rho(1700)$     & 1720 & 250  &1720 & 250   &0.22$\pm$0.03  &8.8$\sigma$                                       \\

\hline
\hline
\end{tabular}
\end{small}
\end{table*}

In the next step,
a search is performed for additional resonances with 
$J^{PC}=$ $1^{-+}_{\eta\etap}, 0^{++}_{\eta\etap}, 2^{++}_{\eta\etap}, 4^{++}_{\eta\etap},$ $1^{+-}_{\gamma\eta^{(\prime)}}$, and $1^{--}_{\gamma\eta^{(\prime)}}$, where the subscript labels the composition of the resonance, by individually adding each possibility to the PDG-optimized solution and scanning  over its mass and width.
The significance of each additional resonance at each mass and width is evaluated. 
The result indicates that a significant $1^{-+}$ contribution ($\textgreater$$7\sigma$) is needed around 1.9 GeV/$c^{2}$ in the $\eta\etap$ system. The significances for all other additional contributions are less than 5$\sigma$.
Therefore, an $\eta_1$ state is included in the PWA.

In the final step,
a baseline set of amplitudes is determined by adding the $\eta_1$ state, with 
its mass and width as free parameters, to the PDG-optimized set of amplitudes. 
The statistical significances of all resonances in the PDG-optimized set are then reevaluated in the presence of the $\eta_1$ state. 
Contributions from the $f_0(2100)$, $h_1(1595)_{\gamma\etap}$, $\rho(1700)_{\gamma\etap}$, $\phi(2170)_{\gamma\eta}$, $f_{2}(1810)$, and $f_{2}(2340)$ in the PDG-optimized set of amplitudes become insignificant ($\textless3\sigma$) and are thus omitted from the baseline set of amplitudes, where the subscript labels the composition of the resonance.
The statistical significance of the $f_4(2050)$ is reduced from 5.6$\sigma$ to 4.6$\sigma$, but is still retained. 
By introducing the $\eta_1$, the mass and width of the $f_0(2020)$ becomes more consistent with the average values in the PDG~\cite{Zyla:2020zbs}. 
In addition, a nonresonant contribution modeled by a $0^{++}$ $\eta\eta^{\prime}$ system uniformly distributed in the phase space, is included with a significance of 15.7 $\sigma$.
After this amplitude selection process, the baseline set of amplitudes includes eleven components. 
The isoscalar state with exotic quantum numbers $J^{PC} = 1^{-+}$, the $\eta_1$, has a mass of (1855$\pm$9$_{\rm {stat}}$)~MeV/$c^{2}$ and a width of (188$\pm$18$_{\rm {stat}}$)~MeV with a statistical significance of 21.4$\sigma$. It is denoted as $\etamp$.

The results of the PWA with the baseline set of amplitudes, including 
the masses and widths of the resonances, the product branching fractions $\BR(\jpsi\rightarrow\gamma X)\BR(X\rightarrow\eta\eta')$ and $\BR(\jpsi\rightarrow\eta' X)\BR(X\rightarrow\gamma\eta)$, and the statistical significances, are summarized in Table~\ref{Summary of all}.
The fit fractions for each component and their interference fractions are shown in Table~\ref{fraction baseline solution}. 
The measured masses and widths of the $f_{0}(2020)$ and $f_{2}(2010)$ are consistent with the PDG~\cite{Zyla:2020zbs} average values. 
The measured mass of the $f_{0}(2330)$, which is unestablished in the PDG~\cite{Zyla:2020zbs}, is consistent with the results of Ref.~\cite{Bugg:2004xu}, but our measured width is 79 MeV smaller (3.4$\sigma$).

All other resonances considered have statistical significance less than 3$\sigma$ when added to the baseline set of amplitudes, 
as shown in Table~\ref{significance check}. 
To investigate additional possible contributions, 
resonances with different $J^{PC}$ ($0^{++}_{\eta\etap},1^{-+}_{\eta\etap}, 2^{++}_{\eta\etap}, 4^{++}_{\eta\etap}, 1^{+-}_{\gamma\eta^{(\prime)}}$, and $1^{--}_{\gamma\eta^{(\prime)}}.$) 
and with different masses and widths are added to the baseline set of amplitudes.
No significant contributions from additional resonances with conventional quantum numbers are found. The most significant additional contribution (4.4$\sigma$) comes from an exotic $1^{-+}$ component around 2.2 GeV. 
Changing the $J^{PC}$ assignment of the $0^{++}$ PHSP component in the baseline set of amplitudes to $1^{-+}_{\eta\etap}, 2^{++}_{\eta\etap}, 4^{++}_{\eta\etap}, 1^{+-}_{\gamma\eta^{(\prime)}}$, and $1^{--}_{\gamma\eta^{(\prime)}}$, 
results in a worse negative log-likelihood by at least 57.
Furthermore, additional nonresonant contributions with all other $J^{PC}$ assignments are found to be insignificant.

\begin{figure*}[htbp]
  \centering
    \subfigure{  	
    \includegraphics[width=0.4\textwidth]{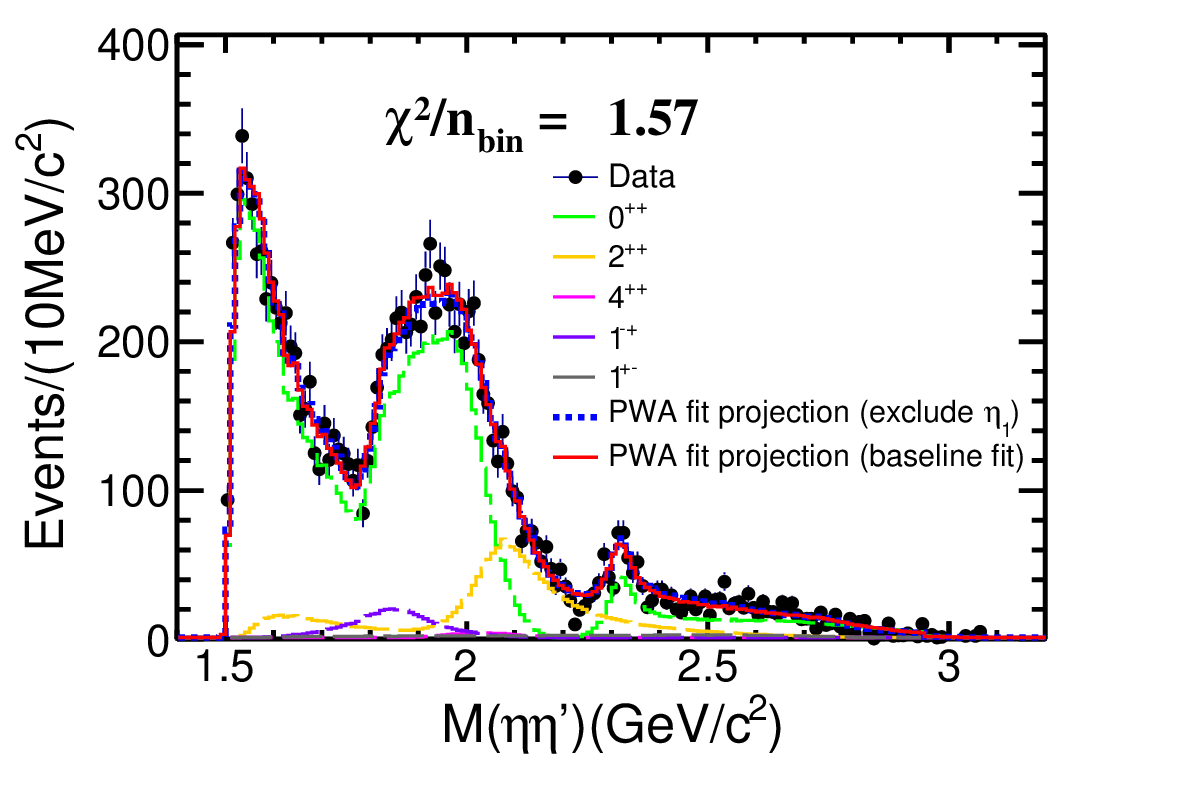}\put(-150,115){(a)}}%
  \subfigure{
    \includegraphics[width=0.4\textwidth]{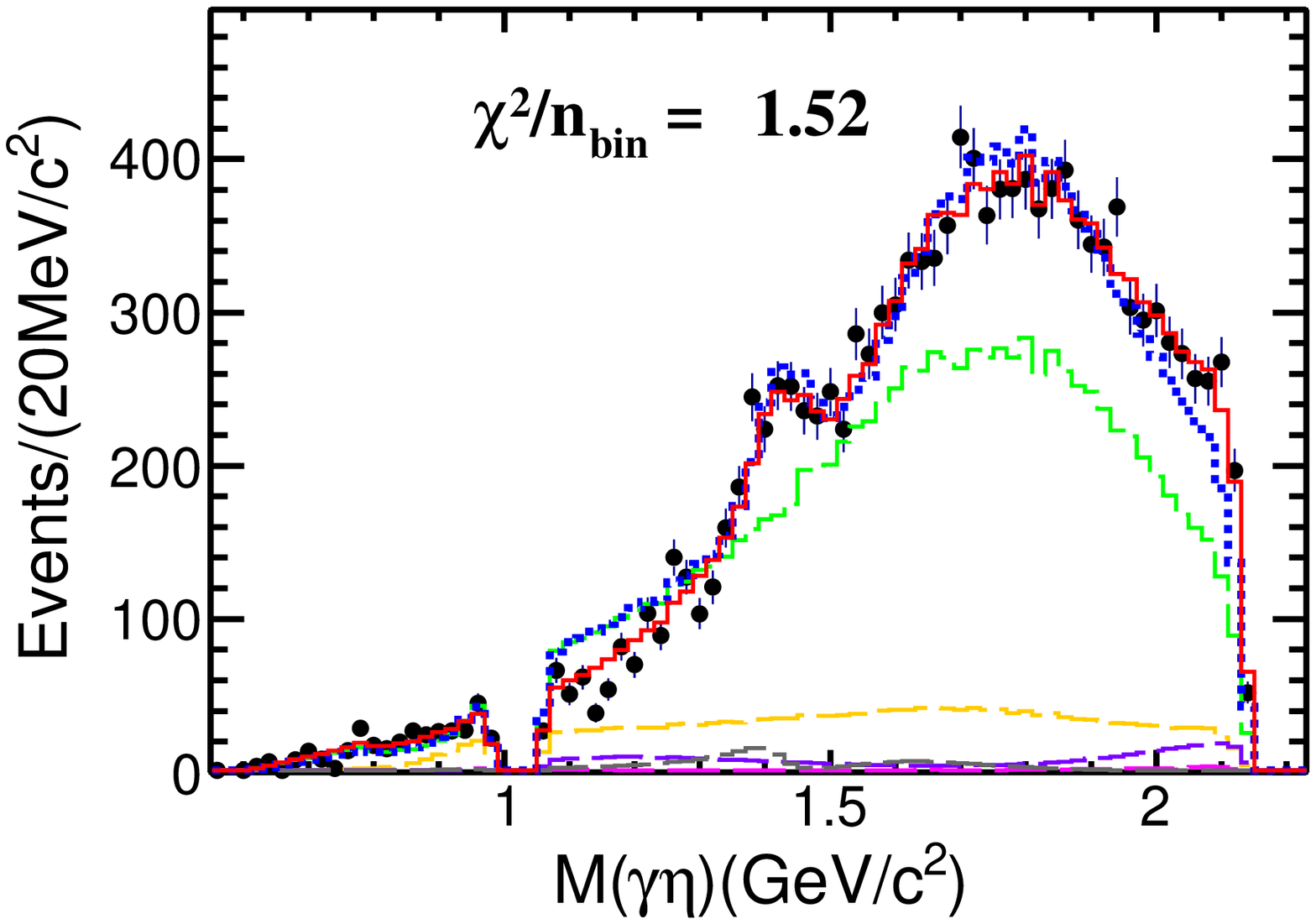}\put(-150,115){(b)}} %
  \subfigure{
    \includegraphics[width=0.4\textwidth]{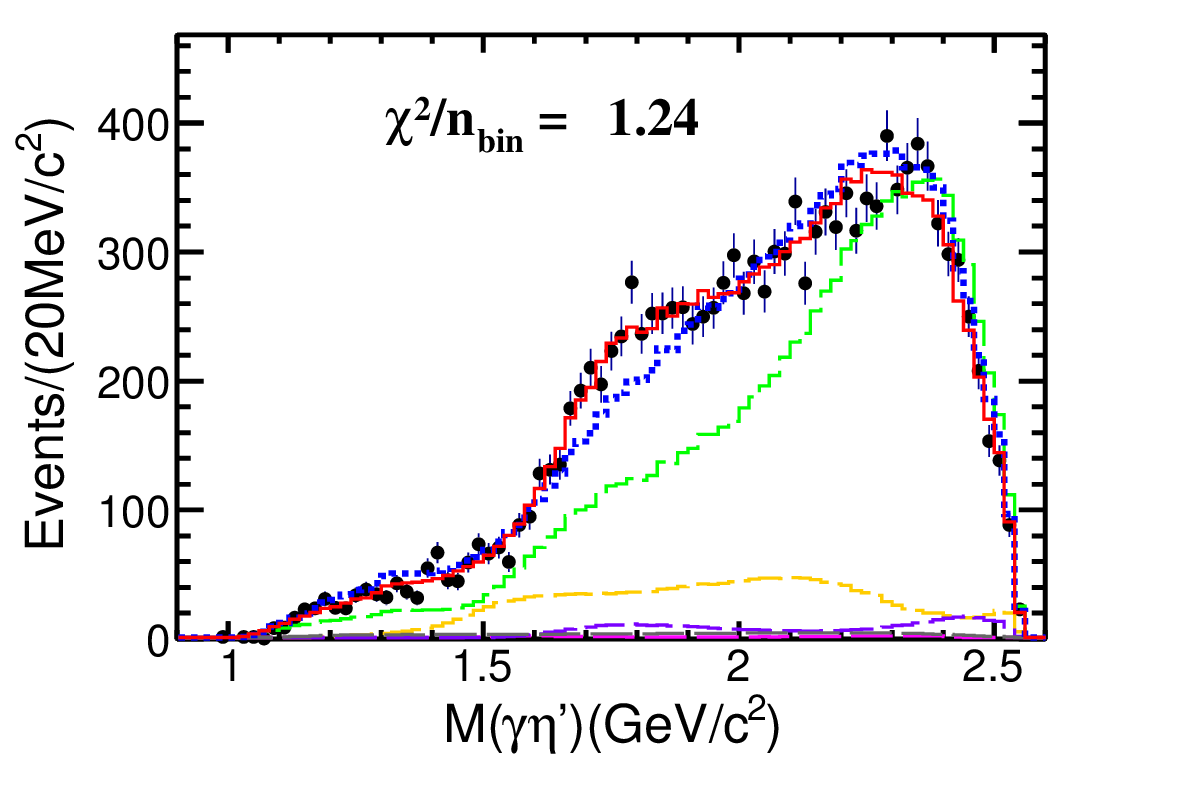}\put(-150,115){(c)}}%
   \subfigure{
     \includegraphics[width=0.4\textwidth]{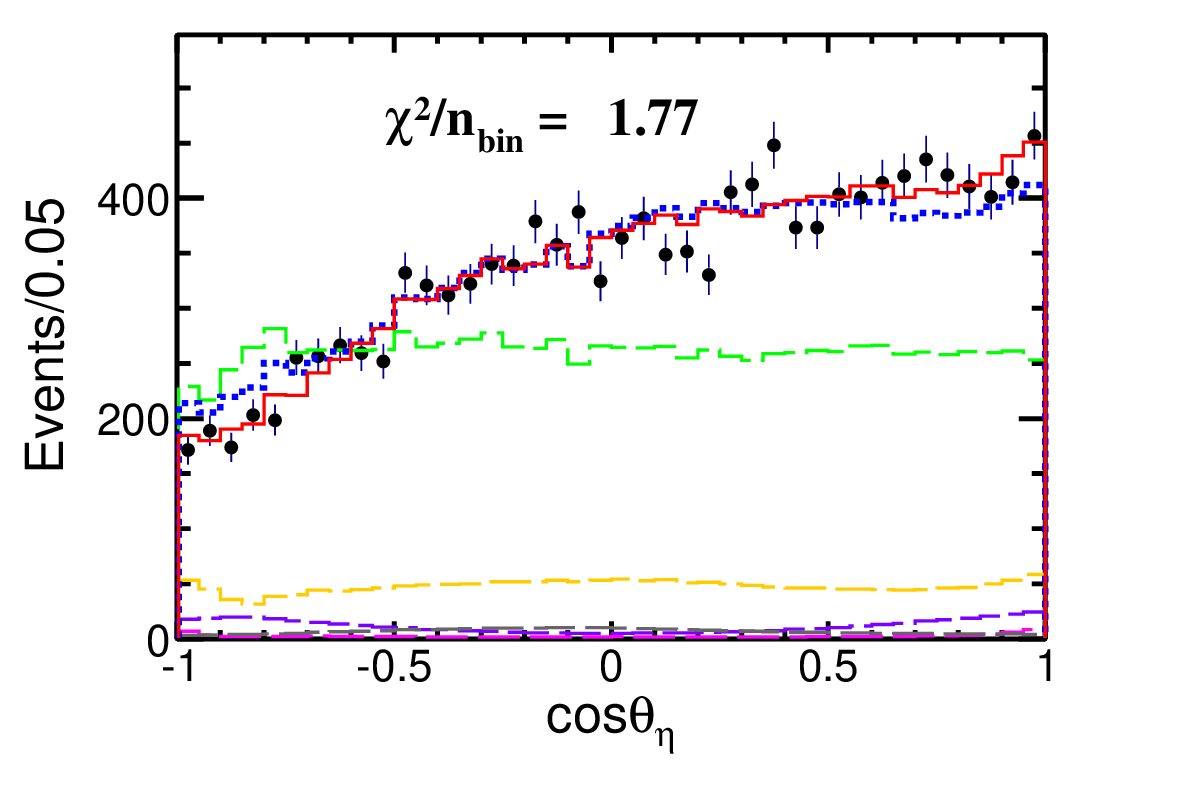}\put(-150,115){(d)}}\\%

%    \subfigure{  	
%    \includegraphics[width=0.4\textwidth]{FullRange_mX.eps}\put(-150,115){(a)}}%
%  \subfigure{
%    \includegraphics[width=0.4\textwidth]{FullRange_mgEta.eps}\put(-150,115){(b)}} %
%  \subfigure{
%    \includegraphics[width=0.4\textwidth]{FullRange_mgEtap.eps}\put(-150,115){(c)}}%
%   \subfigure{
%     \includegraphics[width=0.4\textwidth]{FullRange_ct_eta.eps}\put(-150,115){(d)}}\\%

   \caption{Background-subtracted data (black points) and the PWA fit projections (lines) for (a,b,c) the invariant mass distributions of 
(a) $\eta\etap$, 
(b) $\gamma\eta$,
and (c) $\gamma\eta'$, and (d) the distribution of cos$\theta_{\eta}$,  
where $\theta_{\eta}$ is the angle of the $\eta$ momentum in the $\eta\etap$ helicity coordinate system.
The red lines are the total fit projections from the baseline PWA.  The blue lines are the total fit projections from a fit excluding the $\eta_1$ component. The dashed lines for the $1^{-+}, 0^{++}, 2^{++}, 4^{++}$ and $1^{+-}$ contributions are the coherent sums of amplitudes for each $J^{PC}$.}
 \label{PWA fit plot}
\end{figure*}

\begin{figure*}[htbp]
   \centering
       \subfigure{
          \includegraphics[width=0.32\textwidth]{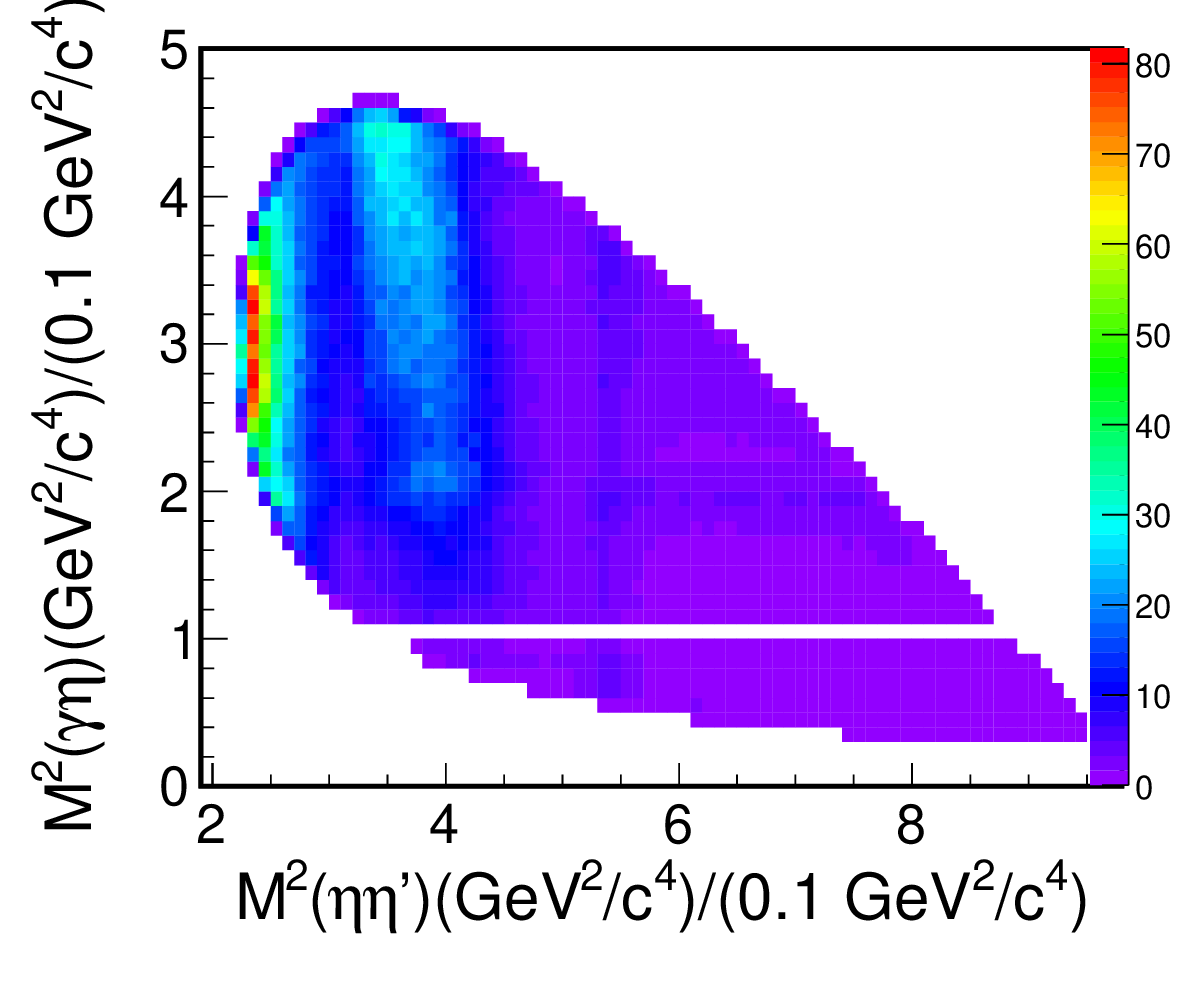}\put(-45,120){(a)}}
         \subfigure{
     \includegraphics[width=0.32\textwidth]{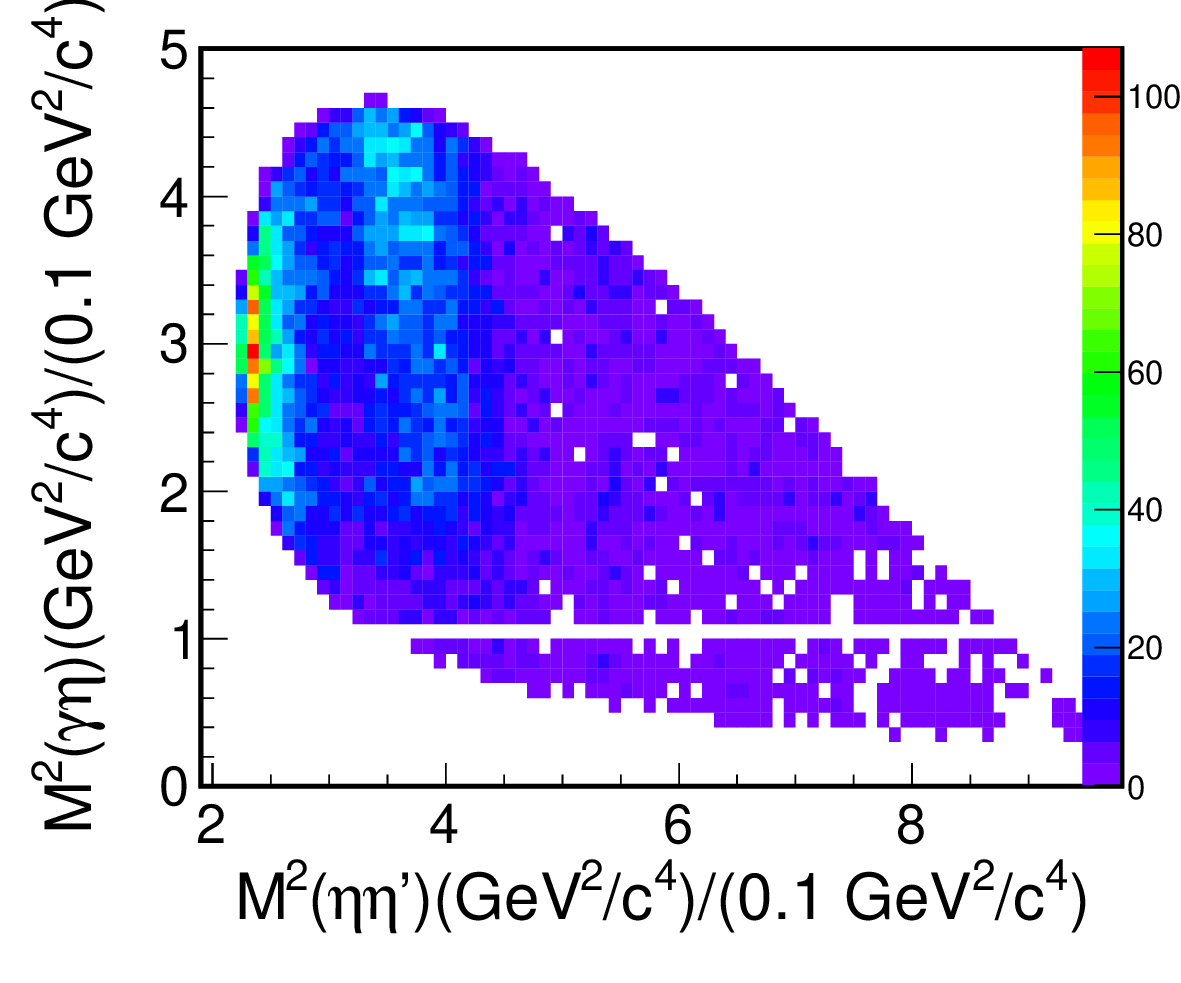}\put(-45,120){(b)}}
          \subfigure{
     \includegraphics[width=0.32\textwidth]{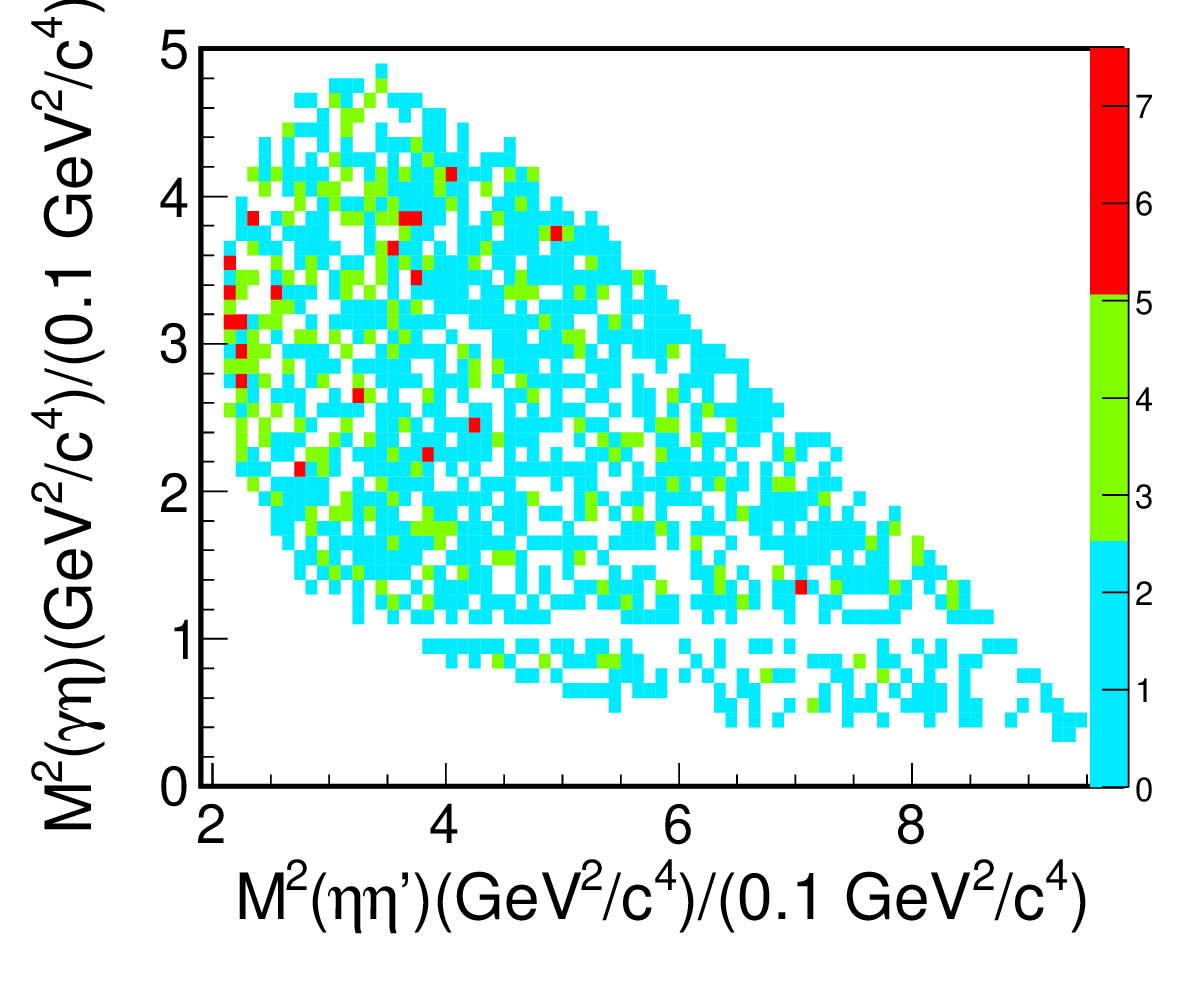}\put(-45,120){(c)}}\\
      \caption{Dalitz plots for  (a) the baseline PWA, (b) the selected data, and (c) background estimated from the $\etap$ sideband.}
  \label{DalitzPlot}
  \end{figure*}

Figure~\ref{PWA fit plot} (a), (b) and (c) show the  invariant mass distributions of $M(\eta\etap)$, $M(\gamma\eta)$, and $M(\gamma\etap)$ for the data (with background subtracted) and the PWA fit projections,  respectively.  
Figure~\ref{PWA fit plot} (d) shows the cos$\theta_{\eta}$ distribution, where $\theta_{\eta}$ is the angle of the $\eta$ momentum in the $\eta\etap$ (Jacob and Wick) helicity frame 
(in which the $\eta\etap$ system is at rest and the z-axis is defined by the momentum of the photon)~\cite{Jacob:1959at}.
This angle carries information about the spin of the particle decaying to $\eta\etap$.
The $\chi^{2}/$n$_{\rm bin}$ value is displayed on each figure to demonstrate the goodness of fit, where n$_{\rm bin}$ is the number of bins in each histogram, and $\chi^{2}$ is defined as:

{\setlength\abovedisplayskip{1pt}
\setlength\belowdisplayskip{1pt}
\begin{eqnarray}\label{chisq/dof}
\chi^2 = \sum_{i=1}^{n_{\rm bin}} \frac{(n_i - v_i)^2}{v_i},
\end{eqnarray}}where $n_i$ and $v_i$ are the number of events for the data and the fit projections with the baseline set of amplitudes in the $i$th bin of each figure, respectively.
In comparison, the $\chi^{2}/$n$_{\rm bin}$ values for the PDG-optimized set of amplitudes for the distributions of $M(\eta\etap)$, $M(\gamma\eta)$, $M(\gamma\etap)$, and cos$\theta_{\eta}$ are  0.26, 0.43, 0.12 and 0.30 worse than baseline set of amplitudes, respectively.
Figure~\ref{DalitzPlot} shows the Dalitz plots for the PWA fit projection from the baseline set of amplitudes, the selected data, and the background estimated from the $\etap$ sideband. Figure~\ref{PWA fit plot} and Figure~\ref{DalitzPlot} indicates that the data and the PWA fit result (baseline set of amplitudes) are in good agreements.
Compared with the PDG-optimized set, the negative log-likelihood value of the baseline set is improved by 32 units and the number of free parameters is reduced by 16.

\begin{table*}[htbp]
\linespread{1.5}
\centering
\begin{small}
\caption{The mass ($M$), width ($\Gamma$), PDG mass ($M_{\rm PDG}$), PDG width ($\Gamma_{\rm PDG}$), significance (Sig.) and the product branching fractions $\BR(J/\psi\rightarrow \gamma X)\BR(X\rightarrow\eta\etap)$ and $\BR(J/\psi\rightarrow\etap X)\BR(X\rightarrow\gamma\eta)$
 (B.F.) of each component in the 
PWA fit using the baseline set of amplitudes.
The first uncertainties are statistical and the second are systematic. }\label{Summary of all}
\begin{tabular}{c|ccccccc}
\hline
\hline
Decay mode &Resonance     &$M$ (MeV/$c^{2}$) &$\Gamma$ (MeV) &$M$$_{\rm PDG}$ (MeV/$c^{2}$)  &$\Gamma_{\rm PDG}$ (MeV)   &B.F. ($\times$10$^{-5}$)  &Sig.  \\

\hline
&$f_{0}(1500)$  &     1506        &     112              &1506  &112                     &1.81$\pm0.11_{-0.13}^{+0.19}$   &$\gg$30$\sigma$             \\

&$f_{0}(1810)$  &   1795           &         95          &1795 &95         &0.11$\pm$$0.01_{-0.03}^{+0.04}$    &11.1$\sigma$            \\

&$f_{0}(2020)$  &    2010$\pm$6$_{-4}^{+6}$         &           203$\pm$9$_{-11}^{+13}$        &1992 &442           &     2.28$\pm$0.12$_{-0.20}^{+0.29}$          &24.6$\sigma$     \\

$J/\psi\rightarrow \gamma X \rightarrow \gamma \eta\etap$&$f_{0}(2330)$  &    2312$\pm$7$_{-3}^{+7}$         &                65$\pm$10$_{-12}^{+3}$      &2314 &144     &     0.10$\pm$0.02$_{-0.02}^{+0.01}$        &13.2$\sigma$      \\

&$\etamp$  &     1855$\pm$9$_{-1}^{+6}$        &             188$\pm$18$_{-8}^{+3}$        &- &-               &       0.27$\pm$$0.04 _{-0.04}^{+0.02}$     &21.4$\sigma$       \\

&$f_{2}(1565)$  &     1542                                &               122    &1542 &122                         &           0.32$\pm$0.05$_{-0.02}^{+0.12}$     &8.7$\sigma$      \\

&$f_{2}(2010)$  &     2062$\pm$6$_{-7}^{+10}$       &               165$\pm$17$_{-5}^{+10}$        &2011 &202               &       0.71$\pm$0.06$_{-0.06}^{+0.10}$         &13.4$\sigma$     \\

&$f_{4}(2050)$  &      2018        &   237                &2018 &237                     &         0.06$\pm$0.01$_{-0.01}^{+0.03}$        &4.6$\sigma$   \\

&$0^{++}$ PHSP  &     -         &   -                     &- &-                &         1.44$\pm$0.15$_{-0.20}^{+0.10}$         &15.7$\sigma$\\
\hline
$J/\psi\rightarrow \eta^{\prime} X \rightarrow \gamma \eta\etap$&$h_{1}(1415)$    &       1416       &    90    &1416 &90                                &       0.08$\pm$0.01$_{-0.02}^{+0.01}$          &10.2$\sigma$    \\

&$h_{1}(1595)$    &        1584      &    384    &1584 &384                                &    0.16$\pm$0.02$_{-0.01}^{+0.03}$            &9.9$\sigma$     \\

\hline
\hline
\end{tabular}
\end{small}
\end{table*}

\begin{table*}[!htbp]
\centering
\begin{small}
\caption{The fit fractions for each component and the interference fractions between two components(\%) in the PWA fit with the baseline set of amplitudes. The uncertainties are
statistical only.}\label{fraction baseline solution}
\resizebox{\textwidth}{20mm}{
\begin{tabular}{c|ccccccccccc}
\hline

Resonance &$f_{0}(1500)$ &$f_{0}(1810)$ &$f_{0}(2020)$ &$f_{0}(2330)$  &$h_{1}(1415)$  &$h_{1}(1595)$  &$\etamp$ &$f_{2}(1565)$  &$f_{2}(2010)$   &$f_{4}(2050)$   &$0^{++}$ PHSP\\
\hline
\hline
$f_{0}(1500)$  & 21.9$\pm$1.4   & $-$4.3$\pm$0.4  &  16.2$\pm$0.5 & $-$1.0$\pm$0.1 &  1.6$\pm$0.2 & $-$1.6$\pm$0.9 & 0.2$\pm$0.0 & 0.2$\pm$0.1 & 0.6$\pm$0.1 & 0.0$\pm$0.0 &13.4$\pm$1.1        \\
\hline
$f_{0}(1810)$  & & 1.4$\pm$0.1  & $-$5.6$\pm$0.6 &  0.4$\pm$0.0 & $-$0.1$\pm$0.0 & 0.6$\pm$0.1 & 0.0$\pm$0.0 &  $-$0.2$\pm$0.0 & 0.1$\pm$0.0 &0.0$\pm$0.0 & 2.0$\pm$0.3           \\
\hline
$f_{0}(2020)$    && &29.5$\pm$1.6  & $-$3.7$\pm$0.5   &0.0$\pm$0.2 &  $-$3.6$\pm$0.4 & 0.2$\pm$0.0  &  1.1$\pm$0.1  & 0.1$\pm$ 0.1 & 0.1$\pm$0.0 & $-$15.9$\pm$1.8           \\
\hline
$f_{0}(2330)$    &&& &1.4$\pm$0.2  & 0.1$\pm$0.0  &  0.3$\pm$0.1 & 0.0$\pm$0.0  &$-$0.1$\pm$0.0 &  $-$0.2$\pm$ 0.0 & 0.0$\pm$0.0 & 2.6$\pm$0.3            \\
\hline
$h_{1}(1415)$   &&&&  &1.1$\pm$0.2  &$-$1.1$\pm$0.3 & $-$0.2$\pm$0.1 &   0.1$\pm$0.1 & 0.2$\pm$0.1 & $$0.0$\pm$0.0 &2.3$\pm$0.3               \\
\hline
$h_{1}(1595)$  &&&&&  &2.1$\pm$0.3  & 0.5$\pm$0.1 & $-$0.3$\pm$0.3 & 0.0$\pm$0.2 & 0.1$\pm$0.0 &  2.3$\pm$1.0            \\
\hline
$\etamp$    &&&&&&   & 3.5$\pm$0.5  & 0.0$\pm$0.0 & $-$0.1$\pm$0.0  &  0.0$\pm$0.0  & 0.1$\pm$0.0        \\
\hline
$f_{2}(1565)$  &&&&&&&    &4.6$\pm$0.7  &$-$0.6$\pm$0.8 &0.0$\pm$0.0 &$-$0.9$\pm$0.1              \\
\hline
$f_{2}(2010)$  &&&&&&&&   &10.2$\pm$0.8  & $-$0.1$\pm$0.1 & 0.2$\pm$0.1               \\
\hline
$f_{4}(2050)$   &&&&&&&&&    &0.8$\pm$0.2  & $$0.0$\pm$0.0 \\
\hline
$0^{++}$ PHSP  &&&&&&&&&&   &18.5$\pm$1.9                 \\

\hline
\hline
\end{tabular}}
\end{small}
\end{table*}

\begin{table}[!htbp]
\centering
\begin{small}
\caption{Additional resonances considered, their $J^{PC}$, 
the change in negative log-likelihood ($\Delta {\rm ln}\mathcal{L}$) when each is added to the baseline set of amplitudes,
the change in the number of free parameters ($\Delta$dof),
and the resulting statistical significance (Sig.).}\label{significance check}
\begin{tabular}{c|ccccc}
\hline
\hline
Decay mode&Resonance      &$J^{PC}$       &$\Delta {\rm ln}\mathcal{L}$       &$\Delta$dof                                     &Sig.  \\
\hline
&$f_{2}(1525)$    & $2^{++}$      &  6.3 &    6                                & 1.9$\sigma$\\
&$f_{2}(1810)$    & $2^{++}$         &  2.7 &   6                                   & 0.7$\sigma$\\
&$f_{0}(1710)$    & $0^{++}$         &  3.4 &     2                                 &2.1$\sigma$\\
&$f_{2}(1910)$    & $2^{++}$         &  3.9 &      6                                &1.1$\sigma$\\
&$f_{2}(1950)$    & $2^{++}$         &  2.6 &       6                               &0.6$\sigma$\\
&$f_{0}(2100)$    & $0^{++}$          & 1.1  &       2                              &1.1$\sigma$\\
&$f_{2}(2150)$     & $2^{++}$         &  2.3 &      6                             &0.5$\sigma$\\
 $J/\psi\rightarrow \gamma X \rightarrow \gamma \eta\etap$&$f_{0}(2200)$     & $0^{++}$          &  0.4 &       2                             &0.4$\sigma$\\
&$f_{2}(2220)$     & $2^{++}$        &  8.6 &           6                           & 2.6$\sigma$\\
&$f_{2}(2300)$     & $2^{++}$         & 7.2  &            6                         & 2.2$\sigma$\\
&$f_{4}(2300)$     & $4^{++}$         &  2.3 &             6                       &0.5$\sigma$\\
&$f_{0}(2330)$     & $0^{++}$         & 1.5  &              2                       &1.2$\sigma$\\
&$f_{2}(2340)$     & $2^{++}$         &  6.3 &              6                       &1.9$\sigma$\\
&$f_{0}(2102)$\cite{Bugg:2004xu}     & $0^{++}$            &  0.1 &                  2                 & 0.2$\sigma$\\
&$f_{2}(2240)$\cite{Bugg:2004xu}     & $2^{++}$            & 2.9 &               6                   &0.7$\sigma$\\
&$f_{2}(2293)$\cite{Bugg:2004xu}     & $2^{++}$           & 4.1  &                  6                 &1.2$\sigma$\\
&$f_{4}(2283)$\cite{Bugg:2004xu}      & $4^{++}$           & 0.9  &                  6                 & 0.1$\sigma$\\

\hline

&$\rho(1450)$    & $1^{--}$           & 3.4  &                 2                   & 2.1$\sigma$\\
&$\rho(1700)$     & $1^{--}$          & 0.8  &                 2                   & 0.7$\sigma$\\
&$\rho(1900)$     & $1^{--}$          & 0.0  &                 2                   & 0$\sigma$\\
$J/\psi\rightarrow \eta^{\prime} X \rightarrow \gamma \eta\etap$&$\omega(1420)$    & $1^{--}$           & 5.3  &                 2                   &2.8$\sigma$\\
&$\omega(1650)$     & $1^{--}$          & 2.6  &                 2                   & 1.7$\sigma$\\
&$\phi(1680)$       & $1^{--}$        &  4.3 &                 2                   & 2.5$\sigma$\\
&$\phi(2170)$       & $1^{--}$        &  0.4 &                 2                   & 0.4$\sigma$\\

\hline

&$h_{1}(1415)$     & $1^{+-}$          & 1.3  &                 4                   &0.5$\sigma$\\
&$h_{1}(1595)$     & $1^{+-}$          &  8.1 &                 4                   &2.9$\sigma$\\
&$\rho(1450)$       & $1^{--}$        &  1.3 &                 2                   & 1.1$\sigma$\\
&$\rho(1700)$       & $1^{--}$        &  3.1 &                 2                   & 2.0$\sigma$\\
$J/\psi\rightarrow \eta X \rightarrow \gamma \eta\etap$&$\rho(1900)$      & $1^{--}$         & 6.1  &                 2                   &3.0$\sigma$\\
&$\omega(1420)$     & $1^{--}$          &  2.5 &                 2                   & 1.7$\sigma$\\
&$\omega(1650)$      & $1^{--}$         &0.8  &                 2                   &0.7$\sigma$\\
&$\phi(1680)$        & $1^{--}$       &   2.1&                 2                   &1.5$\sigma$\\
&$\phi(2170)$         & $1^{--}$      &  0.1 &                 2                   &0.1$\sigma$\\

\hline
\hline
\end{tabular}
\end{small}
\end{table}

\begin{boldmath}
\subsection{Further checks on the $\etamp$}
\end{boldmath}
%%% disscisuion of eta1 %%%%

\begin{figure}[htbp]
\centering
\includegraphics[width=0.5\textwidth]{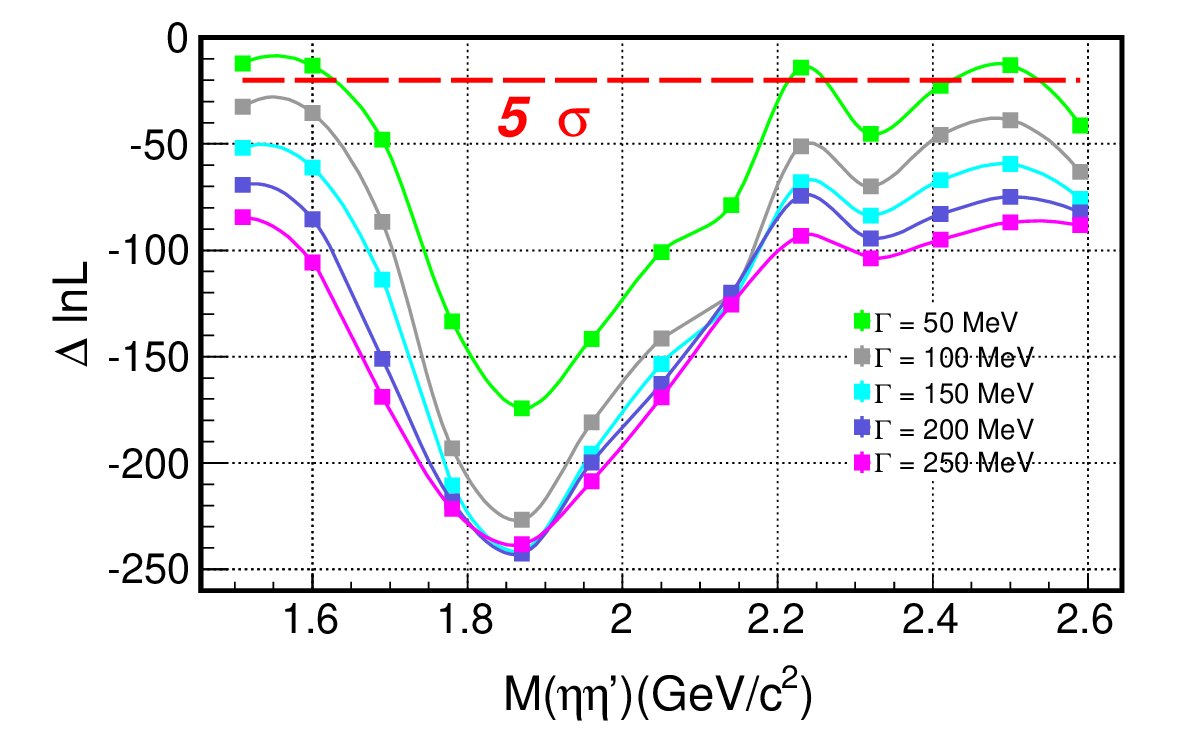}

\caption{The change in negative log-likelihood values for a range of $\eta_{1}$ resonance parameters in the baseline set of amplitudes. The red dashed line indicates that the statistical significances of the hypotheses under the red dashed line are higher than 5$\sigma$. }
\label{Scan eta1 as addition state}
\end{figure}

Various checks are performed to validate the existence of the $\etamp$. 
The fits are carried out by assigning all other possible $J^{PC}$ to the $\eta_{1}(1855)$, and the negative log-likelihoods are worse by at least 235 units (21.8$\sigma$). 
To probe the significance of the BW phase motion, the BW parameterization of the $\etamp$ in the baseline PWA is replaced with an amplitude whose magnitude matches that of a BW function but with constant phase (independent of $s$). This alternative fit has a negative log-likelihood 43 units (9.2$\sigma$) worse than the baseline fit,
which indicates that a resonant structure is favorable. 
In the scenario $\etamp$ is removed from the baseline set of amplitudes, the significance of $\etamp$ with different masses and widths are evaluated.
The changes of negative log-likelihood value are shown in Fig.~$\ref{Scan eta1 as addition state}$. The result shows that a significant $1^{-+}$ contribution is needed around 1.85~GeV/$c^{2}$.

To visualize the agreement between the PWA fit results and data, angular moments as a function of $M(\eta\etap)$ can be calculated for data (with background subtracted) and the PWA model. 
For events within a given region of $M(\eta\etap)$, the cos$\theta_{\eta}$ distribution can be expressed as an expansion in terms of Legendre polynomials. 
The coefficients, 
which are called the unnormalized moments of the expansion, 
characterize the spin of the contributing $\eta\etap$ resonances. 
The moment for the $k$th bin of $M(\eta\etap)$ is
\begin{eqnarray}\label{eqnY}
\langle Y^{0}_{l} \rangle \equiv \sum\limits_{i=1}^{N_{k}}W_{i}Y^{0}_{l}({\rm cos\theta}_{\eta}^{i}).
\end{eqnarray}
For data, $N_{k}$ is the number of observed events in the $k$th bin of $M(\eta\etap)$ and $W_{i}$ is a weight used to implement background subtraction. 
For the PWA model, $N_{k}$ is the number of events in a PHSP MC sample, which is generated with signal events distributed uniformly in phase space,
    and $W_{i}$ is the intensity for each event calculated in the PWA model. 

\begin{figure*}[htbp]
\centering
 \subfigure{
 \label{Y00}
\includegraphics[width=0.38\textwidth]{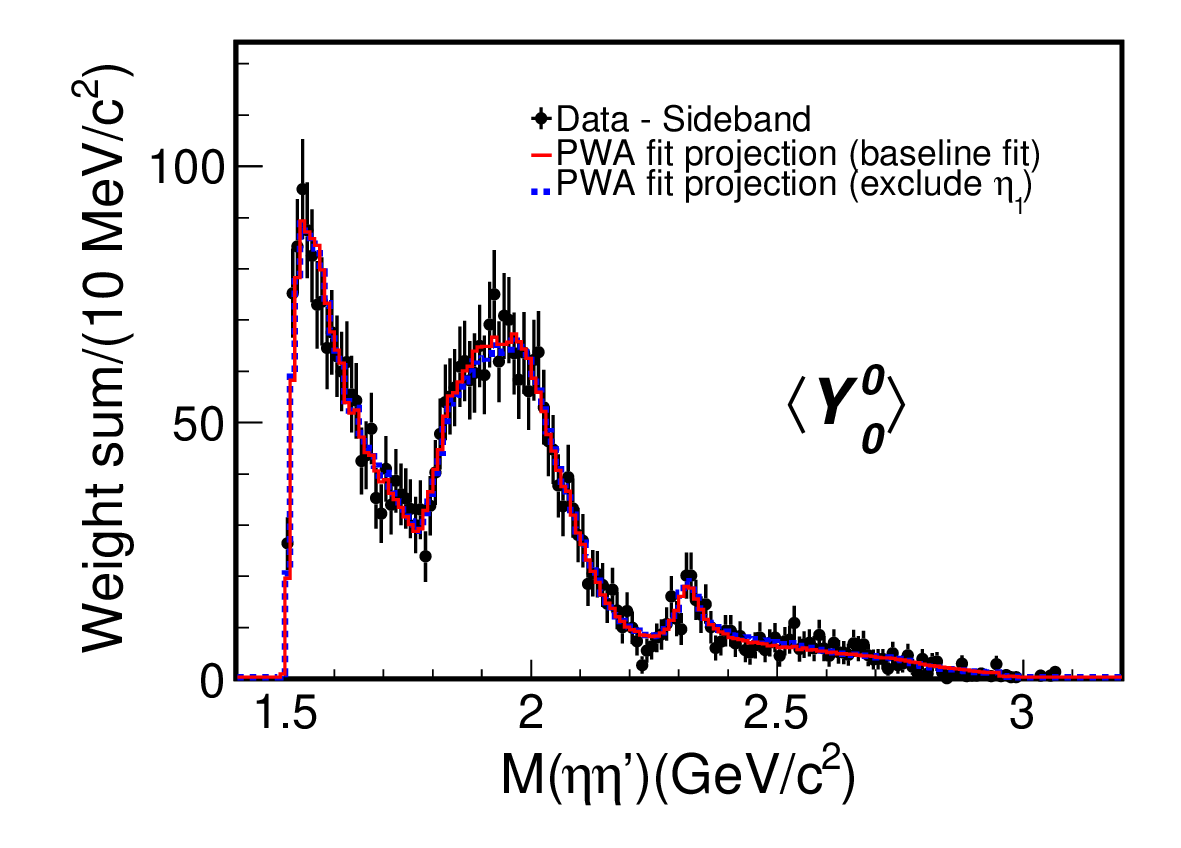}\put(-25,120){(a)}}
\subfigure{
\label{Y01}
\includegraphics[width=0.38\textwidth]{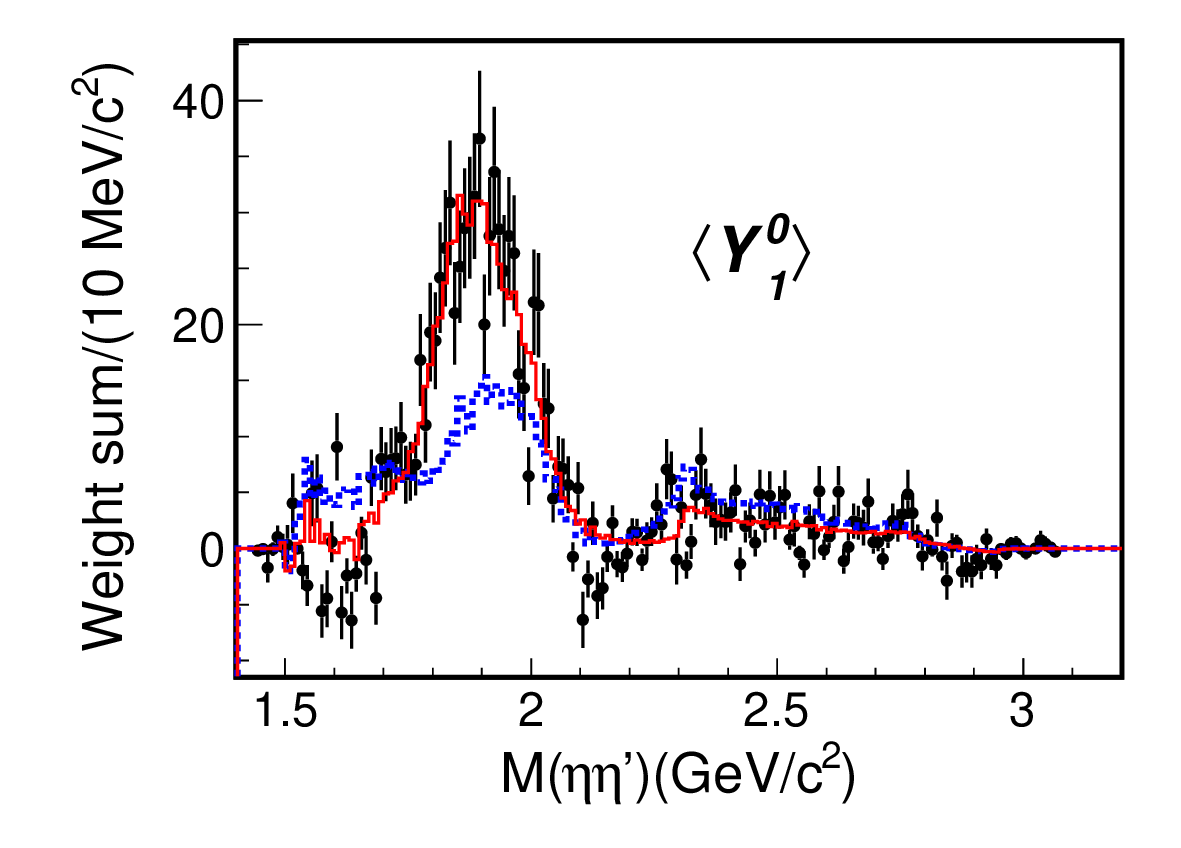}\put(-25,120){(b)}}
\subfigure{
\label{Y02}
\includegraphics[width=0.38\textwidth]{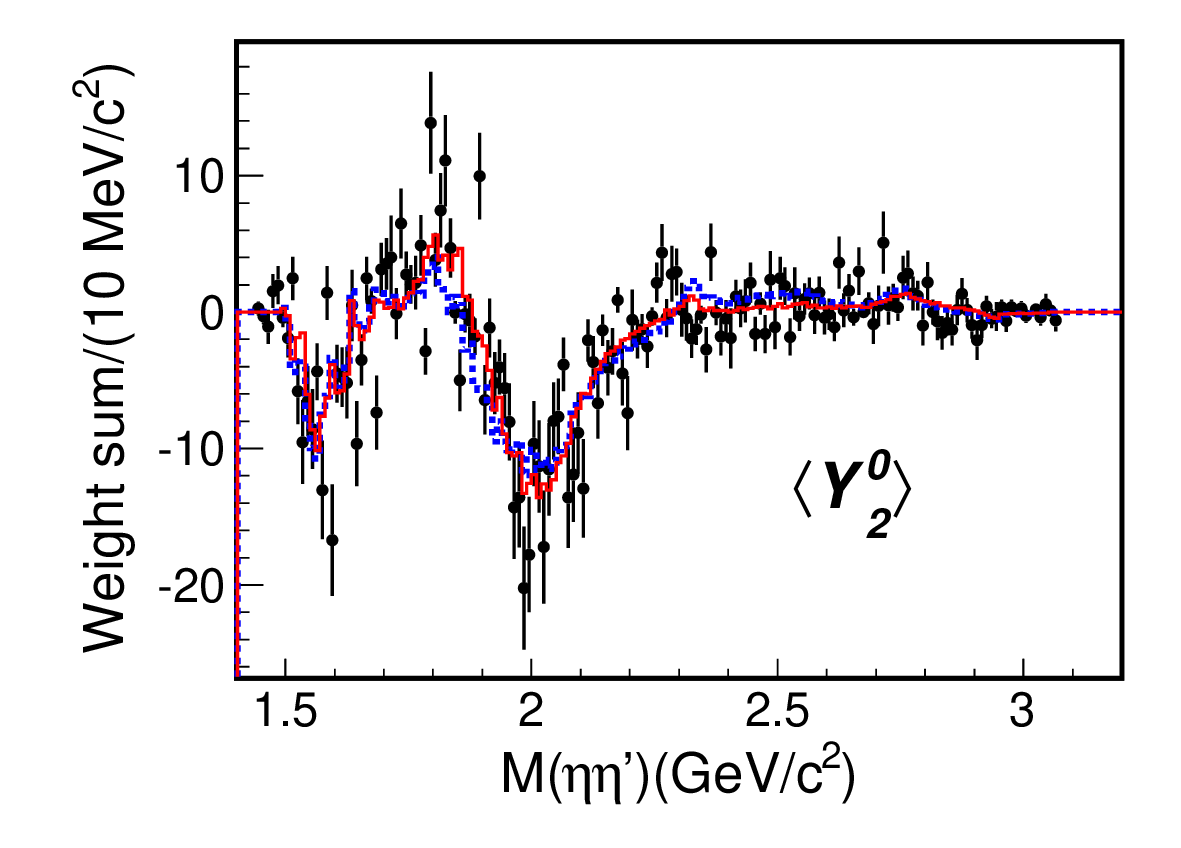}\put(-25,120){(c)}}
\subfigure{
\label{Y03}
\includegraphics[width=0.38\textwidth]{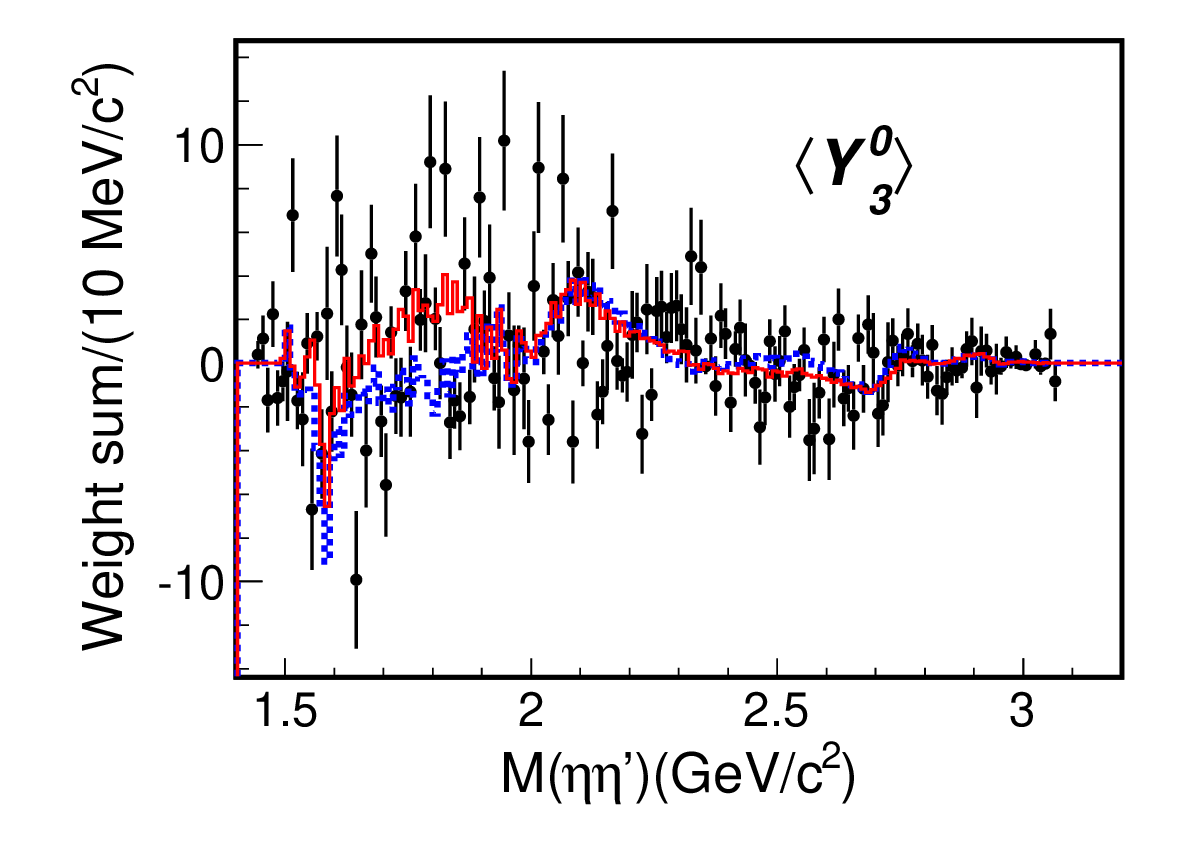}\put(-25,120){(d)}}
\subfigure{
\label{Y04}
\includegraphics[width=0.38\textwidth]{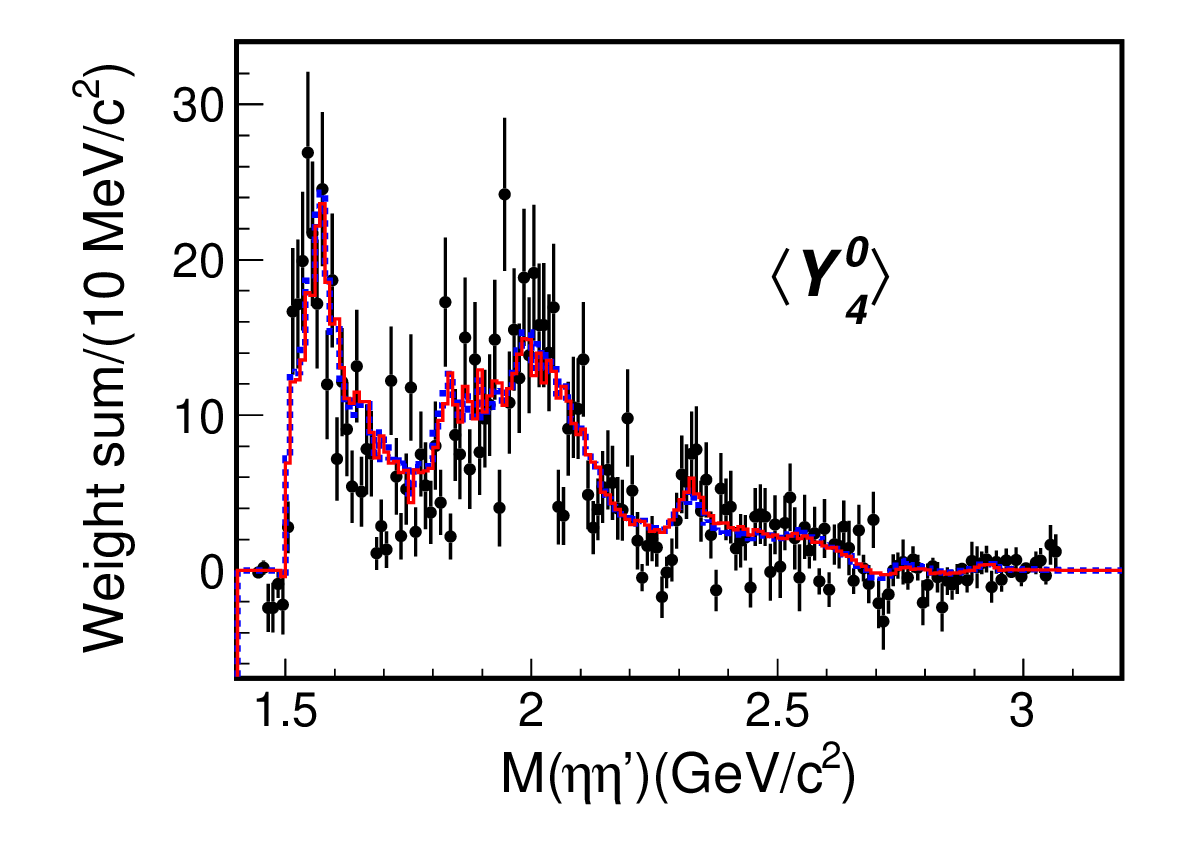}\put(-25,120){(e)}}

% \label{Y00}
%\includegraphics[width=0.37\textwidth]{h0.eps}\put(-25,120){(a)}}
%\subfigure{
%\label{Y01}
%\includegraphics[width=0.37\textwidth]{h1.eps}\put(-25,120){(b)}}
%\subfigure{
%\label{Y02}
%\includegraphics[width=0.37\textwidth]{h2.eps}\put(-25,120){(c)}}
%\subfigure{
%\label{Y04}
%\includegraphics[width=0.37\textwidth]{h4.eps}\put(-25,120){(d)}}

\caption{The distributions of the unnormalized moments 
$\langle Y^{0}_{L} \rangle$ ($L=$ 0, 1, 2, 3 and 4) for $\jpsi\rightarrow\gamma\eta\etap$ as functions of the $\eta\etap$ mass. Black dots with error bars represent the background-subtracted data weighted with angular moments; the red solid lines represent the baseline fit projections; and the blue dotted lines represent the projections from a fit excluding the $\eta_1$ component.}
\label{Angular moments}
\end{figure*}

Neglecting $\eta\eta'$ amplitudes with spin greater than 2, and ignoring the effects of symmetrization and the presence of resonance contributions in the $\gamma\eta$ and $\gamma\eta'$ subsystems, 
the moments are related to the spin-0 ($S$), spin-1 ($P$) and spin-2 ($D$) amplitudes by~\cite{Costa:1980ji,PrivateCommunication}:
\begin{widetext}
\begin{eqnarray}\label{eqnY00}
\sqrt{4\pi}\langle Y^{0}_{0} \rangle = S^{2}_{0} + P^{2}_{0} + P^{2}_{1} + D^{2}_{0} + D^{2}_{1} + D^{2}_{2},
\end{eqnarray}
\begin{eqnarray}\label{eqnY01}
\sqrt{4\pi}\langle Y^{0}_{1} \rangle = 2S_{0}P_{0} \cos\phi_{P_{0}} + \frac{2}{\sqrt{5}}( 2P_{0}D_{0}\cos(\phi_{P_{0}} - \phi_{D_{0}}) + \sqrt{3}P_{1}D_{1}\cos(\phi_{P_{1}} - \phi_{D_{1}} )) ,
\end{eqnarray}
\begin{eqnarray}\label{eqnY02}
\sqrt{4\pi}\langle Y^{0}_{2} \rangle = \frac{1}{7\sqrt{5}}(14P^{2}_{0} - 7P^{2}_{1} + 10D^{2}_{0} + 5D^{2}_{1} - 10D^{2}_{2}) + 2S_{0}D_{0}\cos\phi_{D_{0}}  ,
\end{eqnarray}
\begin{eqnarray}\label{eqnY03}
\sqrt{4\pi}\langle Y^{0}_{3} \rangle = \frac{6}{\sqrt{35}}( \sqrt{3}P_{0}D_{0}\cos(\phi_{P_{0}} - \phi_{D_{0}}) - P_{1}D_{1}\cos(\phi_{P_{1}} - \phi_{D_{1}} ))  ,
\end{eqnarray}
\begin{eqnarray}\label{eqnY04}
\sqrt{4\pi}\langle Y^{0}_{4} \rangle = \frac{1}{7}( 6D^{2}_{0} - 4D^{2}_{1} + D^{2}_{2}),
\end{eqnarray}
\end{widetext}
where $\phi_{P}$ and $\phi_{D}$ are the phases of the P wave and D wave relative to the S wave.
Figure~\ref{Angular moments} shows the moments computed for the data and the PWA model, using Eq.~\ref{eqnY}, where good data/PWA consistency can be seen.  The need for the $\etamp$ P-wave component is apparent in the $\langle Y^{0}_{1} \rangle$ moment [Fig.~\ref{Angular moments}(b)].

\begin{figure*}[htbp]
   \centering
 \iffalse
   		\subfigure{
           \includegraphics[width=0.23\textwidth]{eps/CosTheta_IndifferentRegion/plotFullRegionFittedMass_mX.eps}\put(-90,65){(a)}}
\fi

         \subfigure{
          \includegraphics[width=0.32\textwidth]{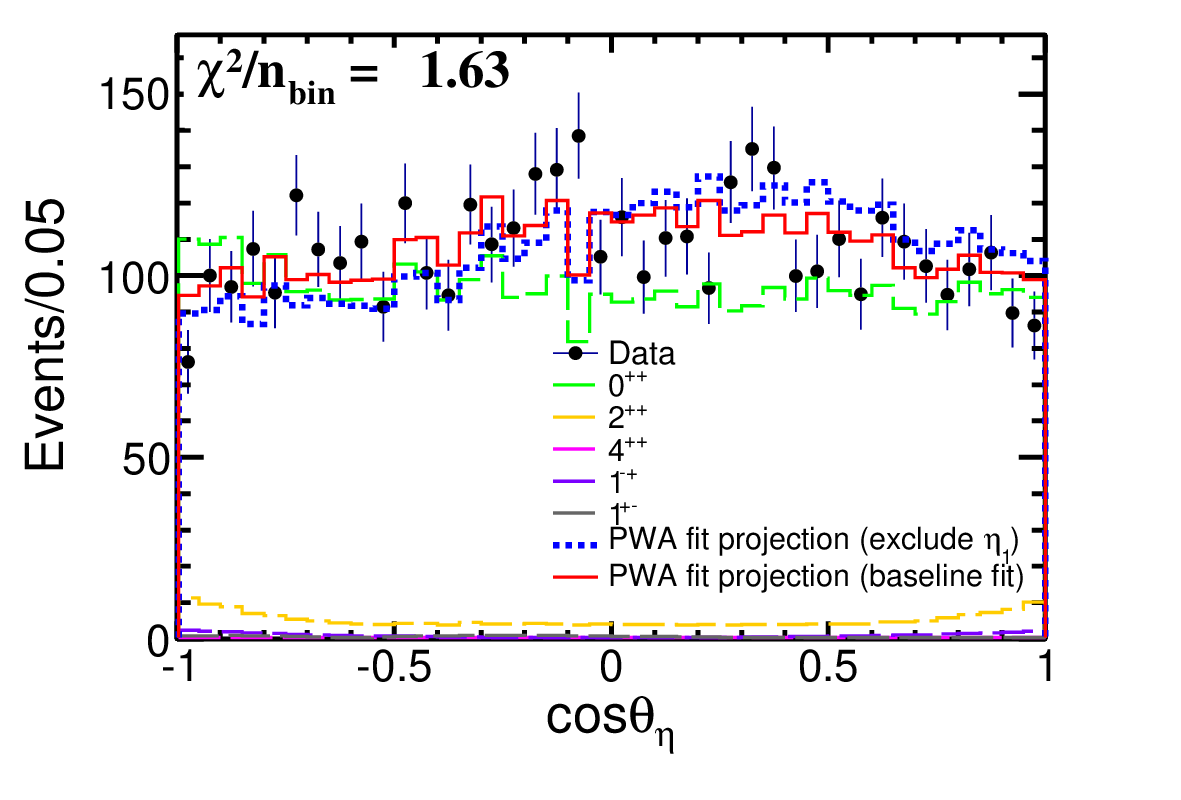}\put(-40,95){(a)}}
         \subfigure{
     \includegraphics[width=0.32\textwidth]{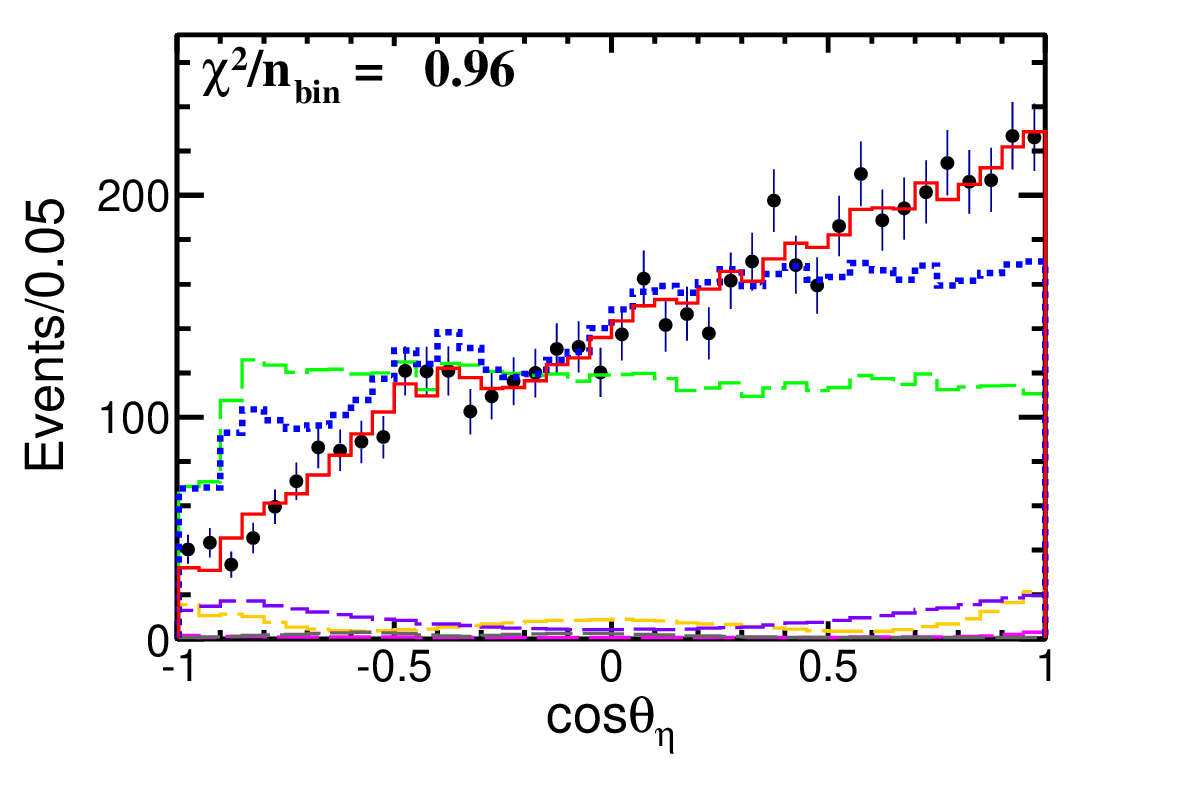}\put(-40,95){(b)}}
          \subfigure{
     \includegraphics[width=0.32\textwidth]{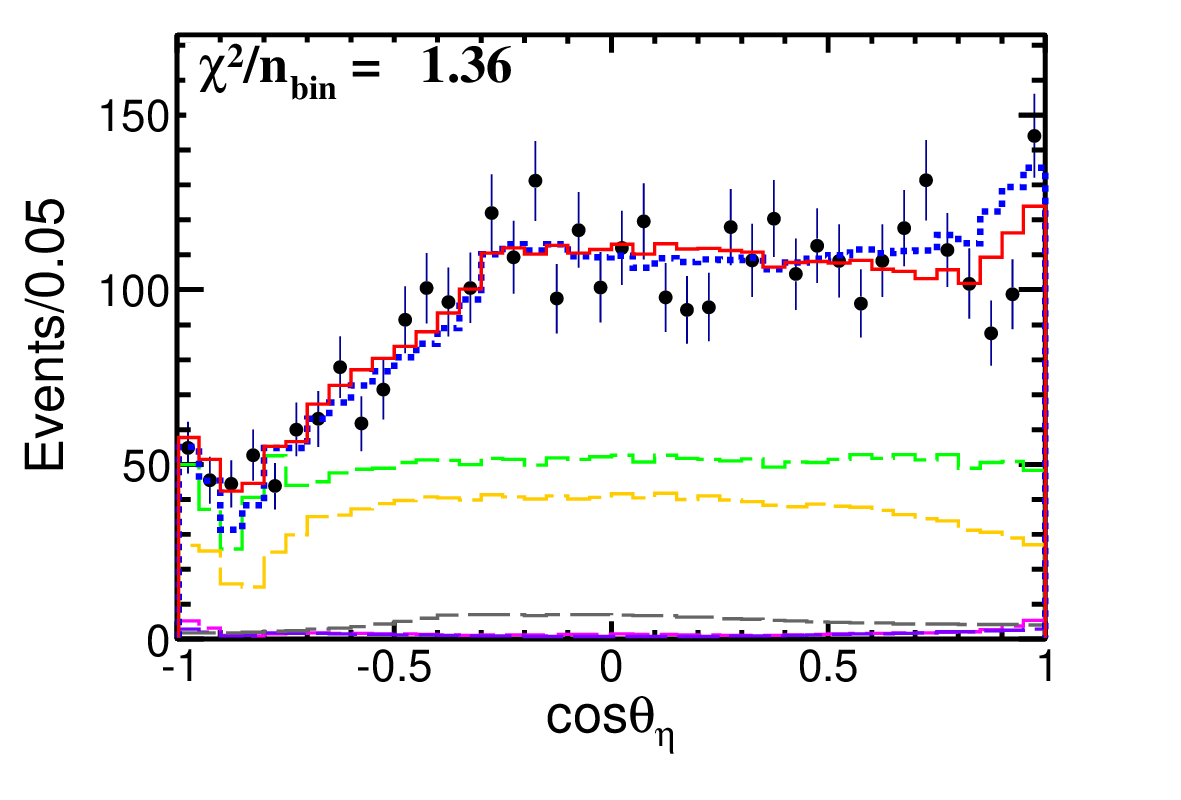}\put(-40,95){(c)}}\\

%         \subfigure{
%          \includegraphics[width=0.32\textwidth]{LowRange_ct_eta.eps}\put(-40,95){(a)}}
%         \subfigure{
%     \includegraphics[width=0.32\textwidth]{MiddleRange_ct_eta.eps}\put(-40,95){(b)}}
%          \subfigure{
%     \includegraphics[width=0.32\textwidth]{HighRange_ct_eta.eps}\put(-40,95){(c)}}\\

   \caption{Background-subtracted data (black points) and the PWA fit projections (lines) for 
the cos$\theta_{\eta}$ distribution when the $\eta\etap$ mass is restricted to the regions:
(a) [1.5,1.7], (b) [1.7,2.0], and (c) [2.0,3.2]~GeV/$c^{2}$.
The red lines are the total fit projections from the baseline PWA.  The blue lines are the total fit projections from a fit excluding the $\eta_1$ component. }
  \label{Angle in different region}
  \end{figure*}

Figure~$\ref{Angle in different region}$ shows a comparison of the data and the PWA projection of cos$\theta_{\eta}$ in different $M(\eta\etap)$ regions ([1.5,1.7], [1.7,2.0] and [2.0,3.2]~GeV/$c^{2}$). 
There is a clear asymmetry in the cos$\theta_{\eta}$ distribution in the 
region [1.7,2.0]~GeV/$c^{2}$ largely due to the $\etamp$ signal, and the $\chi^{2}/$n$_{\rm bin}$ of this region indicates a good agreement between data and the fitting results.

\begin{boldmath}
\subsection{Discussion of the $f_0(1500)$ and $f_0(1710)$}\label{Discuss1710}
\end{boldmath}

The dominant contributions in the baseline PWA are from the scalar resonances. 
A significant signal for the $f_{0}(1500)$ is observed with a large product branching fraction ($\BR(J/\psi$ $\rightarrow$ $\gamma f_{0}(1500))$ $\BR(f_{0}(1500)\rightarrow$$\eta\etap)$
 = (1.81$\pm0.11$$_{\rm {stat}}$)$\times 10^{-5}$). 
Since the mass of the $f_0(1500)$ is close to the $\eta\etap$ mass threshold and the $f_0(1500)$ has other decay modes, we parameterize the $f_0(1500)$ with a ${\rm {Flatt\acute{e}}}$-like form with its mass and width as free parameters. The ${\rm {Flatt\acute{e}}}$-like propagator is 
\begin{eqnarray}\label{BWMD}
f^{(f_0(1500))}_{{\rm {Flatt\acute{e}}}} = \frac{1}{M^2 - s -\frac{1}{c^{2}}i\sqrt{s}\Gamma(s)},
\end{eqnarray}

\begin{eqnarray}\label{BWMDWith}
\Gamma(s) =g\Gamma(\frac{M^2}{s})(\frac{\rho(s)}{\rho(M^2)})^{2l+1} + (1-g)\Gamma_{0},
\end{eqnarray}
where the $\Gamma$ in the first term of $\Gamma(s)$
is an effective parameter corresponds to the decay mode $f_0(1500)\rightarrow\eta\etap$, $l$ is orbital angular momentum of $\eta\etap$ system, $g$ is $\BR(f_{0}(1500) \rightarrow \eta\eta' )$, which is estimated to be 0.02 from the PDG~\cite{Zyla:2020zbs}, the second term corresponds to all other decay modes of the $f_0(1500)$, and $\Gamma_{0}$ is a constant which represents the total width of the $f_0(1500)$ listed in  the PDG~\cite{Zyla:2020zbs}. $\rho(s)$ is the momentum magnitude of $\eta^{(\prime)}$ in the $\eta\etap$ rest frame :
\begin{eqnarray}\label{momentum magnitude}
\rho(s) = \sqrt{\frac{((s+s_{\eta}-s_{\etap})^2}{4s}-s_{\etap}}.
\end{eqnarray}
The impact of using the ${\rm {Flatt\acute{e}}}$-like parametrization for the $f_0(1500)$ is assigned as a systematic uncertainty, which is discussed further in Sec.~\ref{SysUn}. 

The ratio $\BR(f_0$$\rightarrow$$\eta\etap)$/$\BR(f_0$$\rightarrow$$\pi\pi)$ can be calculated with the branching fractions measured in this analysis and in PDG~\cite{Zyla:2020zbs}. 
The ratio $\BR(f_0(1500)$$\rightarrow$$\eta\etap)$/$\BR(f_0(1500)$$\rightarrow$$\pi\pi)$ is determined to be (1.66$\pm0.38_{\rm stat}$)$\times$10$^{-1}$, where the uncertainty is statistical only.
In comparison,  the product branching fraction $\BR(J/\psi$$\rightarrow$$\gamma$$f_{0}(1810))$$\BR(f_{0}(1810)\rightarrow\eta\etap)$
 is (0.11$\pm$$0.01$$_{\rm {stat}}$)$\times10^{-5}$. 
If we consider the $f_{0}(1810)$ and $f_{0}(1710)$ to be the same state, the ratio $\BR(f_0(1810)$$\rightarrow$$\eta\etap)$/$\BR(f_0(1710)$$\rightarrow$$\pi\pi)$ is 
(2.9$^{+1.1}_{-0.9}$)$\times$10$^{-3}$, where the error includes both the systematic and statistical uncertainties.
If the $f_{0}(1710)$ is added to the baseline set of amplitudes, 
the statistical significance of the $f_0(1710)$ is 2.1$\sigma$,
the $\BR(J/\psi$$\rightarrow$$\gamma$$f_{0}(1710))$$\BR(f_{0}(1710)$$\rightarrow$$\eta$$\etap)$
is (1.87$\pm$0.54$_{\rm {stat}}$) $\times$10$^{-7}$, and the $\BR(J/\psi$$\rightarrow$$\gamma$$f_{0}(1500))$$\BR(f_{0}(1500)$$\rightarrow$$\eta$$\etap)$ becomes (1.98$\pm$0.06$_{\rm {stat}}$)$\times$10$^{-5}$. 
If the $f_0(1810)$ is replaced by the $f_0(1710)$ with mass and width taken from the PDG, the product branching fraction 
$\BR(J/\psi$$\rightarrow$$\gamma$$f_{0}(1710))$$\BR(f_{0}(1710)$$\rightarrow$$\eta$$\etap)$
 becomes (7.16$\pm2.21$$_{\rm {stat}}$)$\times 10^{-7}$ and the negative log-likelihood is worse 
by 29 units~(7.3$\sigma$). 
%The branching fraction of $\BR(\jpsi$$\rightarrow$$\gamma f_{0}(1710)$$\rightarrow$$\gamma\eta\eta')$ $=$ (7.16$\pm2.21$$_{\rm {stat}}$)$\times 10^{-7}$.
There is therefore no significant evidence for  $J/\psi$$\rightarrow$$\gamma f_{0}(1710)\rightarrow\gamma\eta\eta'$.
To determine the upper limits on 
$\BR(J/\psi$$\rightarrow$$\gamma$$f_{0}(1710))$$\BR(f_{0}(1710)$$\rightarrow$$\eta$$\etap)$
for different scenarios, the same approach as that in Ref.~\cite{BESIII:2018rdg} is used. For each alternative fit, 
the upper limit is determined at 90$\%$ of the integral of a Gaussian distribution with mean and width equal to the fitting yield and the statistical uncertainty. 
The maximum value is taken as the upper limit.

%%%%% Systematic uncertainty %%%%%%%%
\section{SYSTEMATIC UNCERTAINTIES}\label{SysUn}
%%%%% Systematic uncertainty  :  Event selection %%%%%%%%

The sources of systematic uncertainty are divided into two categories. The first category concerns the  systematic uncertainties related to event selection, which are applicable to measurements of the branching fractions.  These sources of systematic uncertainties are described below. The second category of  systematic uncertainties concerns the PWA and will be treated later.
\begin{enumerate}[(i)]
\item Pion tracking. The MDC tracking efficiency of charged pions is investigated using a clean control sample of $\jpsi \rightarrow p \bar p \pip\pim$~\cite{Ablikim:2011es}. The difference in tracking efficiency between data and MC simulation is 1$\%$ for each charged pion.
\item Photon detection efficiency. The photon detection efficiency is studied with a clean sample of $\jpsi$ $\rightarrow $$\rho^{0}\pi^{0}$~\cite{Ablikim:2010zn}. The result shows that the data-MC efficiency difference is 1$\%$ per photon.
\item  Kinematic fit. 	To investigate the systematic uncertainty associate with the kinematic fit, the track helix parameter correction method~\cite{bib27} is used. The difference in the detection efficiency between using and not using the helix correction is taken as the systematic uncertainty.
\item $\etap$ mass resolution. The difference in the mass resolution between data and MC simulation leads to uncertainties related to the $\etap$ mass window requirement. This is investigated by smearing the MC simulation to improve the consistency between data and MC simulation. The difference of the detection efficiency before and after smearing is assigned as the systematic uncertainty for the $\etap$ mass window requirement.

\item Other systematic uncertainties. The systematic uncertainty due to the number of $J/\psi$ events is determined to be 0.43$\%$ according to Ref.~\cite{Ablikim:NumOfJpsi}. The uncertainties on the intermediate decay branching fractions of $\etap$$\rightarrow$$\eta\pimp$, $\etap$$\rightarrow$$\gamma\pimp$, and $\eta$$\rightarrow$$\gamma\gamma$ are taken from the world average values~\cite{Zyla:2020zbs}, 
which are 0.5$\%$, 0.4$\%$ and 0.2$\%$, respectively.

\end{enumerate}

For the two $\etap$ decay modes, the systematic uncertainties from pion tracking, four photon detection, number of $\jpsi$ events, $\BR(\eta \rightarrow \gamma\gamma$) are common systematic uncertainties and the other systematic uncertainties are independent systematic uncertainties. The combination of common and independent systematic uncertainties for these two decay modes are calculated with the weighted least squares method~\cite{DAgostini:1993arp}, and the total systematic uncertainty is determined to be 4.8$\%$. A summary of all systematic uncertainties related to event selection is shown in Table~\ref{SysEvent}.

\begin{table}[!htbp]
\centering
\begin{small}
\caption{Relative systematic uncertainties related to event selection for the determination of the branching fraction (in $\%$).}\label{SysEvent}
\begin{tabular}{c|c|c}
\hline

\hline
\hline
\multicolumn{3}{c}{Common systematic uncertainties } \\
 \hline
Sources  & $\etap \rightarrow \eta \pip \pim $  & $\etap \rightarrow \gamma \pip \pim$\\
\hline
Pion tracking        &   \multicolumn{2}{c}{2} \\
Four photon detection      & \multicolumn{2}{c}{4} \\
Number of $\jpsi$ events        & \multicolumn{2}{c}{0.43}   \\
$\BR$($\eta\rightarrow \gamma\gamma$)& \multicolumn{2}{c}{0.2}  \\
\hline
Total  & \multicolumn{2}{c}{4.5}   \\

\hline
\hline
 \multicolumn{3}{c}{Independent systematic uncertainties } \\
 \hline
Sources  & $\etap \rightarrow \eta \pip \pim $  & $\etap \rightarrow \gamma \pip \pim$\\
\hline

Another photon detection      &  1     &   -   \\
Kinematic fit                 & 1.5        & 2.6\\
$\etap$ mass resolution       &  0.3 &  0.2 \\
$\BR$($\etap\rightarrow \eta\pip\pim$)& 0.5 &- \\
$\BR$($\etap\rightarrow \gamma\pip\pim$)&- &0.4  \\
$\BR$($\eta\rightarrow \gamma\gamma$) for another one& 0.2 & - \\
\hline
Total                         &1.9   &2.6   \\
\hline
\hline
Combined result  &\multicolumn{2}{c}{4.8} \\
\hline
\end{tabular}
\end{small}
\end{table}

Systematic uncertainties from the PWA impact the branching fractions and resonance parameters.  These are studied below, and the statistical significance of the $\etamp$ is recalculated in every variation.

\begin{enumerate}[(i)]
\item BW parametrization. Uncertainty from the BW parametrization is estimated by the changes in the fit result caused by replacing the constant width $\Gamma_{0}$ of the BW for the threshold state $f_{0}(1500)$ with a mass dependent width as described in Sec.~\ref{Discuss1710}. The statistical significance of the $\etamp$ in this case is 21.8$\sigma$.

\item Uncertainty  from resonance parameters. In the baseline fit, the resonance parameters of the $f_{0}(1500)$, $f_{2}(1565)$, $f_{4}(2050)$, $h_{1}(1415)_{\gamma\eta}$, and $h_{1}(1595)_{\gamma\eta}$ are fixed to PDG~\cite{Zyla:2020zbs} average values, and the resonance parameters of $f_0(1810)$ are fixed to the previous measurment~\cite{BESIII:2012rtd}.
An alternative fit is performed where resonance parameters are allowed to vary within
one standard deviation of the PDG values~\cite{Zyla:2020zbs} and Ref.~\cite{BESIII:2012rtd}, and the changes in the results are taken as systematic uncertainties. 
The statistical significance of the $\etamp$ in this case is 20.6$\sigma$.

\item Background uncertainty. 
To estimate the uncertainty due to the background estimation,
alternative fits are performed using different $\etap$ sideband regions and different background normalization factors. In detail, the background normalization factors are varied by one standard deviation,  which is determined from the fit results of  Fig.~\ref{final etaetap and Dalitz}(a) and Fig.~\ref{final etaetap and Dalitz}(b).
The statistical significance of the $\etamp$ is always above 21.1$\sigma$.
The changes in the branching fractions and resonance parameters are assigned as systematic uncertainties.

\item Uncertainty from additional resonances. 
Uncertainties arising from possible additional resonances are estimated by adding the $f_0(1710)$, $f_2(2220)$, $f_4(2300)$, $h_1(1595)_{\gamma\eta'}$, and $\rho(1900)_{\gamma\eta'}$, which are the most significant additional resonances for each possible $J^{PC}$, into the baseline fit individually. 
The resulting changes in the measurements are assigned as systematic uncertainties. 
With the addition of the $f_0(1710)$, $f_2(2220)$, $f_4(2300)$, $h_1(1595)_{\gamma\eta'}$, and $\rho(1900)_{\gamma\eta'}$, the statistical significances of the $\etamp$ are 22.1$\sigma$, 21.2$\sigma$, 21.1$\sigma$, 19.0$\sigma$, and 21.1$\sigma$, respectively.
\end{enumerate}

For each alternative fit performed to estimate the systematic uncertainties in the PWA fit procedure, 
the changes of the measurements are taken as the one-sided systematic uncertainties. 
For each measurement, the individual uncertainties are assumed to be independent and are added in quadrature to obtain the total systematic uncertainty on the negative and positive sides, respectively. 
The sources of systematic uncertainties affecting 
     the measurements of masses and widths of the $f_{0}(2020)$, $f_0(2330)$, $\etamp$, $f_{2}(2010)$ and their contributions are summarized in Table~\ref{summary of Sys err MW}. The relative systematic uncertainties relevant to the branching fraction measurements are summarized in Table~\ref{Sum of Sys Br}. 

Including the systematic uncertainties, the ratio $\BR(f_0(1500)$$\rightarrow$$\eta\etap)$/$\BR(f_0(1500)$$\rightarrow$$\pi\pi)$ is measured to be (1.66$^{+0.42}_{-0.40}$)$\times$10$^{-1}$, where the error includes both the systematic and statistical uncertainties. To include the systematic uncertainties in the upper limit $\BR(f_0(1710)$$\rightarrow$$\eta\etap)$/$\BR(f_0(1710)$$\rightarrow$$\pi\pi)$, 
the additive systematic uncertainties, i.e., the systematic uncertainties on the upper limit associated with the PWA,
are considered by performing alternative fits and taking the maximum value as the upper limit. The multiplicative systematic uncertainties, i.e. the other systematic uncertainties, are taken into account by dividing by the factor ($1-\delta_{\rm combined}$), where $\delta_{\rm combined}$ is the systematic uncertainties associated with the event selection and the uncertainty of branching fractions and the uncertainty of $J/\psi$$\rightarrow$$\gamma f_{0}(1710)$$\rightarrow$$\gamma\pi\pi$~\cite{Zyla:2020zbs}.
The upper limit on  $\BR(f_0(1710)$$\rightarrow$$\eta\etap)$/$\BR(f_0(1710)$$\rightarrow$$\pi\pi)$ at 90$\%$ C.L. is determined to be 2.87$\times$10$^{-3}$.

\begin{table}[htbp]
\linespread{1.5}
\centering
\begin{small}
\caption{
Systematic uncertainties on the 
masses (in~MeV/$c^{2}$) and widths (in~MeV) of the 
$f_{0}(2020)$, $f_{0}(2330)$, $\etamp$, and $f_{2}(2010)$, denoted as $\Delta M$ and $\Delta \Gamma$, respectively.}\label{summary of Sys err MW}
\begin{tabular}{c|cc|cc|cc|cc}
\hline\hline
\multicolumn{1}{c|}{\multirow{2}{2cm}{Sources}} &\multicolumn{2}{c|}{$f_{0}(2020)$} &\multicolumn{2}{c|}{$f_{0}(2330)$} &\multicolumn{2}{c|}{$\etamp$} &\multicolumn{2}{c}{$f_{2}(2010)$}  \\
                         &$\Delta$$M$ &$\Delta \Gamma$ &$\Delta$$M$ &$\Delta \Gamma$ &$\Delta$$M$ &$\Delta \Gamma$ &$\Delta$$M$ &$\Delta \Gamma$ \\
\hline
Breit-Wigner formula      &$_{-1}^{+0}$  &$_{-0}^{+10}$    &$_{-1}^{+0}$   &$_{-0}^{+1}$   &$_{-1}^{+0}$ &$_{-0}^{+2}$    &$_{-4}^{+0}$  &$_{-0}^{+3}$  \\
\hline
Resonance parameters  &$_{-0}^{+1}$  &$_{-10}^{+0}$    &$_{-3}^{+0}$  &$_{-0}^{+2}$    &$_{-0}^{+2}$  &$_{-1}^{+0}$    &$_{-0}^{+0}$ &$_{-2}^{+0}$    \\
\hline
Extra resonances    &$_{-2}^{+4}$ &$_{-2}^{+9}$   &$ _{-0}^{+7}$&$_{-9}^{+1}$   &$_{-0}^{+4}$ &$_{-6}^{+1}$    &$_{-5}^{+10}$ &$_{-0}^{+10}$ \\
\hline
Background uncertainty  &$_{-1}^{+0}$  &$_{-4}^{+0}$      &$_{-0}^{+3}$  & $_{-7}^{+1}$     &$_{-0}^{+3}$  & $_{-5}^{+1}$       &$_{-1}^{+0}$   &$_{-5}^{+0}$   \\
\hline
Total                &$_{-3}^{+4}$ & $_{-11}^{+13}$   &$_{-3}^{+7}$ &$_{-12}^{+3}$    &$_{-1}^{+6}$ &$_{-8}^{+3}$     &$_{-7}^{+10}$ &$_{-5}^{+10}$ \\
\hline\hline
\end{tabular}
\end{small}
\end{table}

\begin{table*}[htbp]
\linespread{1.5}
\centering
\begin{small}
\caption{
Relative systematic uncertainties on the branching fractions of $\jpsi\rightarrow\gamma X\rightarrow \gamma\eta\eta'$ or $\jpsi\rightarrow\eta^{(\prime)}X\rightarrow\gamma\eta\eta'$ (relative uncertainties, in $\%$).}\label{Sum of Sys Br}
\resizebox{\textwidth}{20mm}{
\begin{tabular}{cccccccccccc}
\hline
Sources     &$f_{0}(1500)$ &$f_{0}(1810)$ &$f_{0}(2020)$ &$f_{0}(2330)$  &$\etamp$  &$f_{2}(1565)$  &$f_{2}(2010)$  &$f_{4}(2050)$ &0$^{++}$PHSP &$h_{1}(1415)(\gamma\eta)$ &$h_{1}(1595)(\gamma\eta)$ \\
\hline
Event selection   &\multicolumn{11}{c}{$_{-4.8}^{+4.8}$}\\
\hline
Breit-Wigner formula  &$_{-1.7}^{+0}$   &$_{-0}^{+11.6}$ &$_{-0}^{+6.9}$  &$_{-0}^{+3.2}$ & $_{-1.1}^{+0}$  &$_{-0}^{+17.8}$ &$_{-0}^{+0.2}$  &$_{-0}^{+4.2}$  &$_{-0.6}^{+0}$  &$_{-8.2}^{+0}$  &$_{-4.1}^{+0}$        \\
\hline
Extra resonances   &$_{-1.0}^{+9.4}$ &$_{-8.4}^{+30.4}$   &$_{-0}^{+10.0}$  &$_{-13.4 }^{+7.8 }$  &$_{-10.4 }^{+3.5 }$ &$_{-2.7}^{+31.5 }$  & $_{-6.5 }^{+12.9 }$ &$_{-4.7 }^{+44.4 } $ &$_{-12.2 }^{+5.1 }$  &$_{-9.1 }^{+11.0 }$  &$_{-2.2 }^{+16.2 }$      \\
\hline
Resonance parameters   &$_{-4.8}^{+0}$  &$_{-25.6}^{+0}$  &$_{-6.5}^{+0}$  &$_{-0}^{+3.6}$ &$_{-6.1}^{+0}$  &$_{-0}^{+5.5}$  &$_{-0}^{+0.2}$  &$_{-1.4}^{+0}$  &$_{-4.6}^{+0}$  &$_{-11.4}^{+0}$  &$_{-4.3}^{+0}$   \\
\hline
Backgroud uncertainty &$ _{-0.6 }^{+0.5 }$   & $_{-7.5 }^{+0.4  }$  & $_{-3.4  }^{+0.8  }$  & $_{-10.4  }^{+0.3  }$  & $_{-1.1  }^{+0.2  }$  & $_{-0}^{+11.0  } $ &$_{-2.7  }^{+0}  $& $_{-6.5  }^{+31.9  } $ & $_{-1.8  }^{+0} $ &$ _{-8.8  }^{+0}$  & $_{-0.6  }^{+8.4  }$  \\
\hline
Total               &$_{-7.1 }^{+10.6 } $&$_{-28.4 }^{+32.9 }$ & $_{-8.8 }^{+13.1 }$& $_{-17.6 }^{+10.3 }$& $_{-13.1 }^{+5.9 }$&$ _{-5.5 }^{+38.5 }$& $_{-8.5 }^{+13.7 }$& $_{-9.5 }^{+55.0 }$& $_{-14.0 }^{+7.0 }$& $_{-19.5 }^{+12.0 }$& $_{-8.0 }^{+18.8 }$ \\
\hline\hline
\end{tabular}}
\end{small}
\end{table*}
%%%%%%%%%  Summary   %%%%%%%%%%%%%%%%
\section{Summary}

In summary, a PWA of $J/\psi\rightarrow\gamma\eta\etap$ has been performed based on (10.09$\pm$0.04)$\times$10$^{9}$ $\jpsi$ events collected with the BESIII detector.  An isoscalar state with exotic quantum numbers $J^{PC}=1^{-+}$, denoted as $\etamp$, has been observed for the first time. 
Its mass and width are measured  to be (1855$\pm$9$_{-1}^{+6}$)~MeV/$c^{2}$ and (188$\pm$18$_{-8}^{+3}$)~MeV, 
which are consistent with LQCD calculations for 
the $1^{-+}$ hybrid~\cite{Dudek:2013yja}. 
The first uncertainties are statistical and the second are systematic. 
The statistical significance of the resonance hypothesis is estimated to be larger than 19$\sigma$. The product  branching fraction $\BR(J/\psi\rightarrow \gamma\eta_1(1855))$$\BR(\etamp\rightarrow\eta\eta')$ is measured to be (2.70$\pm0.41_{-0.35}^{+0.16})\times$10$^{-6}$.
Further study with more production mechanisms and decay modes will help to identify the nature of $\etamp$.
The decay $\jpsi\rightarrow\gamma f_{0}(1500)\rightarrow\gamma\eta\etap$ has also been observed ($\textgreater$30$\sigma$), while  $\jpsi\rightarrow\gamma f_{0}(1710)\rightarrow\gamma\eta\etap$ is found to be insignificant.
The ratio  $\BR(f_0(1500) \rightarrow \eta\etap)$/$\BR (f_0(1500) \rightarrow \pi\pi)$ is  measured to be (1.66$^{+0.42}_{-0.40}$)$\times$10$^{-1}$, which is consistent with the PDG value~\cite{Zyla:2020zbs}.
For the first time, the upper limit on the ratio of  $\BR(f_0(1710) \rightarrow \eta\etap)$/$\BR (f_0(1710) \rightarrow \pi\pi)$ at 90$\%$ confidence level is determined to be 2.87$\times$10$^{-3}$.
The suppressed decay rate of the $f_0(1710)$ into $\eta\etap$ lends further support to the hypothesis that the $f_0(1710)$ has a large overlap with the ground state scalar glueball~\cite{etaetapBR2}, and the $f_0(1710)$/$f_0(2020)$ might be interpreted as flavor singlet~\cite{Klempt:2021wpg}.

%%%%%%%%%  Acknowledgement   %%%%%%%%%%%%%%
\section*{ACKNOWLEDGMENT}
The BESIII collaboration thanks the staff of BEPCII and the IHEP computing center for their strong support. This work is supported in part by National Key R$\&$D Program of China under Contracts Nos. 2020YFA0406300, 2020YFA0406400; National Natural Science Foundation of China (NSFC) under Contracts Nos. 11625523, 11635010, 11675183, 11735014, 11822506, 11835012, 11922511, 11935015, 11935016, 11935018, 11961141012, 12022510, 12025502, 12035009, 12035013, 12061131003; the Chinese Academy of Sciences (CAS) Large-Scale Scientific Facility Program; the CAS Center for Excellence in Particle Physics (CCEPP); Joint Large-Scale Scientific Facility Funds of the NSFC and CAS under Contracts Nos. U1732263, U1832207; CAS Key Research Program of Frontier Sciences under Contract No. QYZDJ-SSW-SLH040; 100 Talents Program of CAS; INPAC and Shanghai Key Laboratory for Particle Physics and Cosmology; ERC under Contract No. 758462; European Union Horizon 2020 research and innovation programme under Contract No. Marie Sklodowska-Curie grant agreement No 894790; German Research Foundation DFG under Contracts Nos. 443159800, Collaborative Research Center CRC 1044, GRK 214; Istituto Nazionale di Fisica Nucleare, Italy; Ministry of Development of Turkey under Contract No. DPT2006K-120470; National Science and Technology fund; Olle Engkvist Foundation under Contract No. 200-0605; STFC (United Kingdom); The Knut and Alice Wallenberg Foundation (Sweden) under Contract No. 2016.0157; The Royal Society, UK under Contracts Nos. DH140054, DH160214; The Swedish Research Council; U. S. Department of Energy under Contracts Nos. DE-FG02-05ER41374, DE-SC-0012069.

\nocite{*}
\bibliographystyle{apsrev4-1}
\bibliography{gammaEtaEtapPRD}

%\begin{thebibliography}{99}
%
%\bibitem{PrivateCommunication} private communication with Alessandro Pilloni
%\end{thebibliography}

\end{document}